\documentclass[11pt,twocolumn]{article}

\usepackage[utf8]{inputenc}
\usepackage{color}
\usepackage{amsmath}
\usepackage{graphicx}
\usepackage{paralist}
\usepackage{url}
\usepackage{makecell}
\usepackage{subcaption}
\usepackage{authblk}
\usepackage{multirow}
\usepackage{amssymb}
\usepackage{amsmath}
\usepackage{amsfonts}
\usepackage{stackengine}
\stackMath

\newtheorem{example}{Example}
\newtheorem{definition}{Definition}

\usepackage{bera}%
\usepackage{listings}
\usepackage{xcolor}

\usepackage{float}

\newfloat{lstfloat}{htbp}{lop}
\floatname{lstfloat}{Listing}

\colorlet{punct}{red!60!black}
\definecolor{background}{HTML}{EEEEEE}
\definecolor{delim}{RGB}{20,105,176}
\colorlet{numb}{magenta!60!black}

\lstdefinelanguage{json}{
    basicstyle=\normalfont\ttfamily,
    string=[s]{"}{"},
    stringstyle=\color{purple!90},
    comment=[l]{:},
    commentstyle=\color{black!90},
    numbers=left,
    numberstyle=\scriptsize,
    stepnumber=1,
    numbersep=8pt,
    showstringspaces=false,
    breaklines=true,
    frame=lines,
    backgroundcolor=\color{background}
}

\newcommand{\alaska}[0]{Alaska}

\newcommand{\myparagraph}[1]{{\vspace{0.5cm} \noindent \textbf{#1.}}}

\newcommand{\sacatalog}[0]{\texttt{Nguyen}}
\newcommand{\flex}[0]{\textsc{FlexMatcher}}
\newcommand{\sources}{\mathcal{S}}

\title{Alaska: A Flexible Benchmark for Data Integration Tasks}

\author[1]{Valter Crescenzi}
\author[1]{Andrea De Angelis}
\author[1]{Donatella Firmani}
\author[1]{Maurizio Mazzei}
\author[1]{Paolo Merialdo}
\author[1]{Federico Piai}
\author[2]{Divesh Srivastava}

\affil[1]{Roma Tre University, \texttt{name.lastname@uniroma3.it}}
\affil[2]{AT\&T Chief Data Office, \texttt{divesh@att.com}}

\date{}

\begin{document}

\maketitle

\begin{abstract}
\sloppy
Data integration is a long-standing interest of the data management community and has many disparate applications, including business, science and government.
We have recently witnessed impressive results in specific data integration tasks, such as Entity Resolution, thanks to the increasing availability of benchmarks. A limitation of such benchmarks is that they typically come with their own task definition and it can be difficult to leverage them for complex integration pipelines. As a result, evaluating end-to-end pipelines for the entire data integration process is still an elusive goal.
In this work, we present \alaska{}, the first benchmark based on real-world dataset to support seamlessly multiple tasks (and their variants) of the data integration pipeline. The dataset consists of ~70k heterogeneous product specifications from 71 e-commerce websites with thousands of different product attributes. Our benchmark comes with profiling meta-data, a set of pre-defined use cases with diverse characteristics, and an extensive manually curated ground truth.
We demonstrate the flexibility of our benchmark by focusing on several variants of two crucial data integration tasks, \emph{Schema Matching} and \emph{Entity Resolution}. Our experiments show that our benchmark enables the evaluation of a variety of methods that previously were difficult to compare, and can foster the design of more holistic data integration solutions. 
\end{abstract}

\section{Introduction}

Data integration is the fundamental problem of providing a unified view over multiple data sources. The sheer number of ways humans represent and misrepresent information makes data integration a challenging problem. Data integration is typically thought of as a \emph{pipeline} of multiple tasks rather than a single problem. Different works identify different tasks as constituting the data integration pipeline, but some tasks are widely recognized at its core, namely, Schema Matching (SM)~\cite{rahm2001survey, bernstein2011generic} and Entity Resolution (ER)~\cite{elmagarmid2006duplicate}. A simple illustrative example is provided in Figure~\ref{fig:tasks} 
involving two sources A and B, with source A providing two camera specifications with two attributes each, and source B providing two camera specifications with three attributes each.  SM results in an integrated schema with four attributes, and ER results in an integrated set of three entities, as shown in the unified view in Figure~\ref{fig:tasks}.

\begin{figure}[h]
\centering
\resizebox{0.9\columnwidth}{!}{%
\begin{tabular}{|c|c|c|} 
\multicolumn{3}{c}{Source A} \\
\hline
ID & Model & Resolution \\ \hline \hline
A.1 & \makecell{Canon \\ 4000D} & 18.0Mp \\
\hline
A.2 & \makecell{Canon \\ 250D} & 24.1Mp \\
\hline
\end{tabular}%
\quad
\begin{tabular}{|c|c|c|c|} 
\multicolumn{4}{c}{Source B} \\
\hline 
ID & Model & Sensor & MP \\ \hline \hline
B.1 & \makecell{Canon \\ 4000D} & CMOS & 17 \\
\hline
B.2 & \makecell{Kodak \\ SP1} & CMOS & 18 \\
\hline
\end{tabular}}

\resizebox{0.8\columnwidth}{!}{%
\begin{tabular}{|c|c|c|c|} 
\multicolumn{4}{c}{} \\
\multicolumn{4}{c}{Unified view} \\
\hline
ID & \textbf{Model} \textit{(A, B)}& \makecell{\textbf{Resolution} \textit{(A)} \\ \textbf{MP} \textit{(B)}}& \textbf{Sensor} \textit{(B)})\\ \hline \hline
\makecell{\textbf{A.1} \\ \textbf{B.1}} & Canon  4000D \textit{(A, B)} & \makecell{18.0Mp \textit{(A)} \\ 17 \textit{(B)}} & CMOS \textit{(B)}\\
\hline
A.2 & \makecell{Canon \\ 250D} \textit{(A)}& 24.1Mp \textit{(A)}& \textit{(null)} \\
\hline
B.2 & \makecell{Kodak \\ SP1} \textit{(B)}& 18 \textit{(B)} & CMOS \textit{(B)}\\
\hline
\end{tabular}}
\caption{Illustrative example of Schema Matching and Entity Resolution tasks. Boldface column names represent attributes matched by Schema Matching, boldface row IDs represent rows matched by Entity Resolution.
} \label{fig:tasks}
\end{figure}

Although fully automated pipelines are still far from being available~\cite{doan2012principles}, we have recently witnessed impressive results in specific tasks (notably, Entity Resolution~\cite{brunner2020entity,cappuzzo2020creating,ebraheem2017deeper,li2020deep,mudgal2018deep}) fueled by recent advances in the fields of machine learning~\cite{goodfellow2016deep} and natural language processing~\cite{devlin2018bert} and to the unprecedented availability of real-world  benchmarks for training and testing (such as, the Magellan data repository~\cite{magellandata}). The same trend -- i.e., benchmarks influencing the design of innovative solutions -- was previously observed in other areas. For instance, the Transaction Processing Performance Council proposed the family of TPC benchmarks\footnote{\url{http://www.tpc.org}} for OLTP and OLAP~\cite{nambiar2006making,poess2000new}, NIST promoted several initiatives with the purpose of supporting research in Information Retrieval tasks such as TREC,\footnote{\url{https://trec.nist.gov}} the ImageNet benchmark~\cite{deng2009imagenet} supports research on visual object recognition.

\sloppy
In the data integration landscape, the Transaction Processing Performance Council has developed TPC-DI~\cite{poess2014tpc}, a data integration benchmark that focuses on the typical ETL tasks, i.e., extraction and transformation of data from a variety of sources and source formats (e.g., CSV, XML, etc.).
Unfortunately, for the ER and SM tasks, only separate task-specific benchmarks are available. For instance, the aforementioned Magellan data repository provides datasets for the ER task where the SM problem has been already solved.
Thus, the techniques influenced by such task-specific benchmarks can be difficult to integrate into a complex data integration pipeline. As a result, the techniques evaluated with such benchmarks are limited by choices made in the construction of the datasets and cannot take into account all the challenges that might arise when tackling the integration pipeline  using the original source data.

\sloppy
\myparagraph{Our approach} Since it is infeasible to get a benchmark for the entire pipeline by just composing isolated benchmarks for different tasks, we provide a benchmark that can support \emph{by design} the development of end-to-end data integration methods. %
Our intuition is to take the best of both real-world benchmarks and synthetic data generators, by providing a real-world benchmark that can be used flexibly to support a variety of data integration tasks and datasets with different characteristics. As for the tasks, we start from the core of the data integration pipeline and focus on popular variants of Schema Matching and Entity Resolution tasks. As for the datasets, we include data sources with different characteristics, yielding different configurations of the dataset (e.g., by selecting sources with longer textual descriptions in attribute values or more specifications). We refer to our benchmark as \alaska{}. %

\sloppy
\myparagraph{The \alaska{} benchmark} The \alaska{} dataset\footnote{\url{https://github.com/merialdo/research.alaska}} consists of almost 70k product specifications extracted from 71 different data sources available on the Web, over 3 domains, or \emph{verticals}: \texttt{camera}, \texttt{monitor} and \texttt{notebook}. Figure~\ref{list:spec_1} shows a sample specification/record from the \texttt{www.buy.net} data source of the \texttt{camera} domain. Each record is extracted from a different web page and refers to a single product. Records contain attributes with the product properties (e.g., the camera resolution), and are represented in a flat JSON format. The default \texttt{<page title>} attribute shows the html title of the original web page and typically contains some of the product properties in an unstructured format. The entire dataset comprises 15k distinct attribute names.
Data has been collected from real-life web data sources by means of a three-steps process: {$(i)$}~web data sources in our benchmark are discovered and crawled by the \texttt{Dexter} focused crawler~\cite{qiu2015dexter}, {$(ii)$}~product specifications are extracted from web pages by means of an ad-hoc method, and {$(iii)$}~ground truth for the benchmark tasks is manually curated by a small set of domain experts, prioritizing annotations with the notion of \emph{benefit} introduced in~\cite{firmani2016online}. %

Preliminary versions of our \alaska{} benchmark has been recently used for the 2020 SIGMOD Programming Contest and for two editions of the DI2KG challenge.%

The main properties of the \alaska{} benchmark are summarized below.

\begin{itemize}
    \item \textbf{Supports Multiple tasks.} \alaska{} supports a variety of data integration tasks. The version described in this work focuses on Schema Matching and Entity Resolution, but our benchmark can be easily extended to support other tasks (e.g., Data Extraction). For this reason, \alaska{} is suitable for holistically assessing complex data integration pipelines, solving more than one individual task. 
    
    \item \textbf{Heterogeneous.} Data sources included in \alaska{} cover a large spectrum of data characteristics, from  small and clean sources to large and dirty ones. Dirty sources may present a variety of records, each providing a different set of attributes, with different representations, while clean sources have more homogeneous records, both in the set of the attributes and in the representation of their values. To this end, \alaska{} comes with a set of profiling metrics that can be used to select subsets of the data, yielding different use-cases, with tunable properties.

    \item \textbf{Manually curated.} \alaska{} comes with a ground truth that has been manually curated by domain experts. Such a ground truth is large both in terms of number of sources and in the number of records and attributes. For this reason, \alaska{} is suitable for evaluating methods with high accuracy, without overlooking tail knowledge.    
\end{itemize}

\myparagraph{Paper outline} Section~\ref{sec:related} discusses related work. Section~\ref{sec:prel} describes the \alaska{} data model and the supported tasks. Section~\ref{sec:profiling} provides a profiling of the datasets and of the ground truth. A pre-defined set of use cases and experiments demonstrating the benefits of using \alaska{} is illustrated in Section~\ref{sec:casestud}).
Section~\ref{sec:coll} describes the data collection process behind the \alaska{} benchmark. Section~\ref{sec:dep} summarizes the experiences with the \alaska{} benchmark in the DI2KG Challenges and in the 2020 SIGMOD Programming Contest. Section~\ref{sec:concl} concludes the paper and briely presents future works.

\section{Related Works}
\label{sec:related}

We now review recent benchmarks for data integration tasks -- namely Schema Matching and Entity Resolution -- either comprising real-world datasets or providing utilities for generating synthetic ones.

\myparagraph{Real-world datasets for SM} In the context of SM, the closest works to ours are XBenchMatch and T2D. XBenchMatch~\cite{duchateau2014designing,duchateau2007xbenchmatch} is a popular benchmark for XML data, comprising data from a variety of domains, such as finance, biology, business and travel. Each domain includes two or a few more schemata. %
T2D~\cite{ritze2015matching} is a more recent benchmark for SM over HTML tables, including hundreds of thousands of tables from the Web Data Commons archive,\footnote{\url{http://webdatacommons.org/}} with DBPedia serving as the target schema. %
Both XBenchMatch and T2D focus on dense datasets, where most of the attributes have non-null values. In contrast, we provide both dense and sparse sources, allowing each of the sources in our benchmark to be selected as the target schema, yielding a variety of SM problems with varying difficulty. Other benchmarks provides smaller datasets and they are suitable either for  a pairwise SM task~\cite{alexe2008stbenchmark,guo2013matchbench}, like in XBenchMatch, or for a many to one SM task~\cite{do2002comparison}, like in T2D. %
Finally, we mention the Ontology Alignment Evaluation Initiative (OAEI)~\cite{algergawy2019results} and \textit{SemTab (Tabular Data to Knowledge Graph Matching)}~\cite{jimenez2020semtab} for the sake of completeness. Such datasets are related to the problem of matching ontologies (OAEI) and HTML tables (SemTab) to relations of a Knowledge Graph and can be potentially used for building more complex datasets for the SM tasks.

\myparagraph{Real-world datasets for ER} There is a wide variety of real-world datasets for the ER task in literature. %
The Cora Citation Matching Dataset~\cite{cora} is a popular collection of bibliography data, consisting of almost 2K citations. %
The Leipzig DB Group Datasets~\cite{rahm,kopcke2010evaluation} provide a collection of bibliography and product datasets that have been obtained by querying  different websites. The Magellan Data Repository~\cite{magellandata} includes a collection of 24 datasets. %
Each dataset represents a domain (e.g., movies and restaurants) and consists of two tables extracted from websites of that domain (e.g., Rotten Tomatoes and IMDB for the movies domain). %
The rise of ER methods based on deep neural networks ~\cite{brunner2020entity,ebraheem2017deeper,li2020deep,mudgal2018deep} and their need for large amounts of training data has recently motivated the development of the Web Data Commons collection (WDC)~\cite{DBLP:conf/www/PrimpeliPB19}, consisting of more than 26M product offers originating from 79K websites. %
Analogously to SM, all the above benchmarks come with their own variant of the ER task. For instance, the Magellan Data Repository provides pairs of sources with aligned schemata and considers the specific problem of finding matches between instances of the two sources; WDC focuses on the problem of matching instances based on the page title or on product identifiers. Differently from the above benchmarks, we consider sources with different data distributions and schemata, yielding a variety of ER problems, with different levels of difficulty. We also note that except for the smallest Cora Dataset, the ground truth of all the mentioned datasets is built automatically (e.g., using a key attribute, such as the UPC code)\footnote{Relying on a product identifier, such as the UPC, can produce incomplete results since there are many standards, and hence the same product can be associated with  different codes in different sources.} with a limited amount of manually verified data ($\approx$2K matching pairs per dataset). In contrast, the entire ground truth available in our \alaska{} benchmark has been manually verified by domain experts.

\myparagraph{Data generators} %
While real-world data have fixed size and characteristics, data generators can be flexibly used in a controlled setting. MatchBench~\cite{guo2013matchbench}, for instance, can generate various SM tasks by injecting different types of synthetic noise on real schemata~\cite{kim1991classifying}. STBenchmark~\cite{alexe2008stbenchmark} provides tools for generating synthetic schemata and instances with complex correspondences.  The Febrl system~\cite{christen2008febrl} includes a data generator for the ER task, that can create fictitious hospital patients data with a variety of cluster size distributions and error probabilities. More recently, EMBench++~\cite{DBLP:conf/iswc/IoannouV14} provides a system to generate data for benchmarking ER techniques, consisting of a collection of matching scenarios, capturing basic real-world situations (e.g., syntactic variations and structural differences) as well as more advanced situations (i.e., evolving information over time). Finally, the iBench~\cite{arocena2015ibench} metadata generator can be used to evaluate a wide-range of integration tasks, including SM and ER, allowing control of the size and characteristics of the data, such as schemata, constraints, and mappings. In our benchmark, we aim to get the best of both worlds: considering real-world datasets and providing more flexibility in terms of task definition and selection of data characteristics.

\section{Benchmark Tasks}
\label{sec:prel}

\begin{figure}
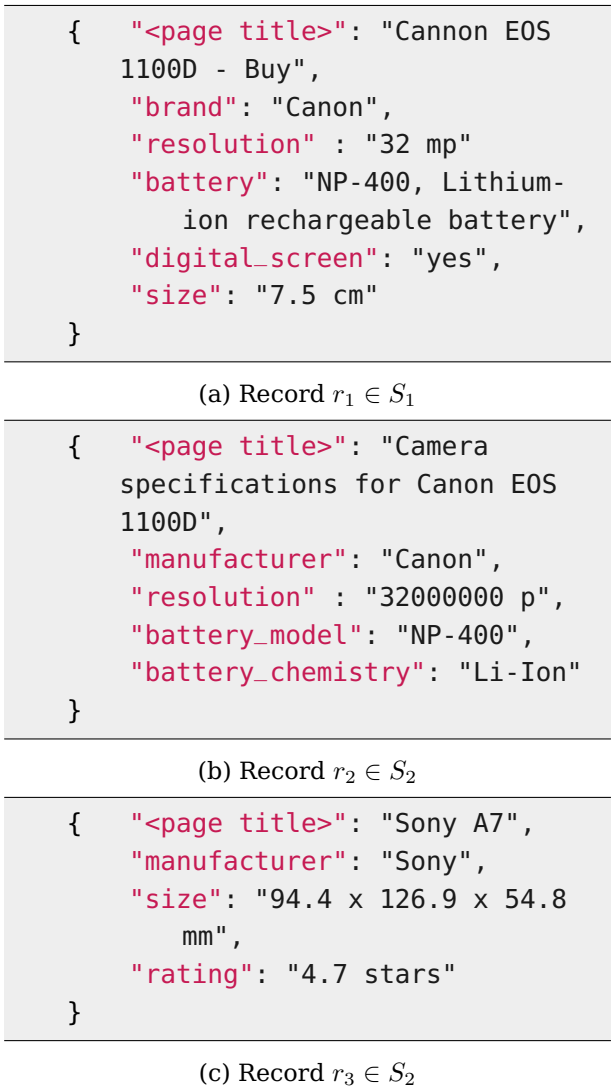

    \centering
    \begin{subfigure}{\columnwidth}
    \begin{lstlisting}[language=json,firstnumber=1]
    {   "<page title>": "Cannon EOS 1100D - Buy",
        "brand": "Canon",
        "resolution" : "32 mp"
        "battery": "NP-400, Lithium-ion rechargeable battery",
        "digital_screen": "yes",
        "size": "7.5 cm"
    }
    \end{lstlisting}
    \vspace{-0.2cm}
    \caption{Record $r_1 \in S_1$\label{list:spec_1}}    
    \vspace{0.2cm}
    \end{subfigure}
    
    \begin{subfigure}{\columnwidth}
    \begin{lstlisting}[language=json,firstnumber=1]
    {   "<page title>": "Camera specifications for Canon EOS 1100D",
        "manufacturer": "Canon",
        "resolution" : "32000000 p",
        "battery_model": "NP-400",
        "battery_chemistry": "Li-Ion"
    }
    \end{lstlisting}
    \vspace{-0.2cm}
    \caption{Record $r_2 \in S_2$\label{list:spec_2}}  
    \vspace{0.2cm}
    \end{subfigure}
    
    \begin{subfigure}{\columnwidth}
    \begin{lstlisting}[language=json,firstnumber=1]
    {   "<page title>": "Sony A7",
        "manufacturer": "Sony",
        "size": "94.4 x 126.9 x 54.8 mm",
        "rating": "4.7 stars"
    }
    \end{lstlisting}
    \vspace{-0.2cm}
    \caption{Record $r_3 \in S_2$\label{list:spec_3}}    
    \end{subfigure}    
    
    \caption{Toy example. \label{fig:list}}
\end{figure}

Let $\sources$ be a set of data sources providing information about a set of \emph{entities}, where each entity is representative of a real-world product (e.g., Canon EOS 1100D). Each source $S \in \sources$ consists of a set of records, where each \emph{record} $r \in S$ refers to an entity and the same entity can be referred to by different records. Each record consists of a title and a set of \emph{attribute-value pairs}. Each attribute refers to one or more \emph{properties}, and the same property can be referred to by different attributes. We describe below a toy example that will be used in the rest of this section.

\begin{example}[Toy example]
Consider the sample records in Figure~\ref{fig:list}. Records $r_1$ and $r_2$ refer to entity ``Canon EOS 1100D'' while $r_3$ refers to ``Sony A7''. Attribute \texttt{brand} of $r_1$ and \texttt{manufacturer} of $r_2$ refer to the same underlying property. 
Attribute \texttt{battery} of $r_1$ refer to the same properties as \texttt{battery\_model} and \texttt{battery\_chemistry} of $r_2$. 
\end{example}

We identify each attribute with the combination of the source name and the attribute name -- such as, $S_1$.\texttt{battery}. We refer to the set of attributes specified for a given record $r$ as $schema(r)$ and to the set of attributes specified for a given source $S \in \sources$, that is, the \emph{source schema}, as $schema(S)=\bigcup_{r \in S} schema(r)$. Note that in each record we only specify the non-null attributes. %

\begin{example}
\sloppy
In our toy example, $schema(S_1)=schema(r_1)=\{S_1$.\texttt{brand}, $S_1$.\texttt{resolution}, $S_1$.\texttt{battery}, $S_1$.\texttt{digital\_screen}, $S_1$.\texttt{size}$\}$, while $schema(S_2)$ is the union of $schema(r_2)$ and $schema(r_3)$, i.e., $schema(S_2)=\{S_2$.\texttt{manufacturer}, $S_2$.\texttt{resolution}, $S_2$.\texttt{battery\_model}, $S_2$.\texttt{battery\_chemistry}, $S_2$.\texttt{size}, $S_2$.\texttt{rating}$\}$.
\end{example}

Finally, we refer to the set of text tokens appearing in a given record $r$ as $tokens(r)$ and to the set of text tokens appearing in a given source, that is, the \emph{source vocabulary}, $S \in \sources$, as $tokens(S)=\bigcup_{r \in S} tokens(r)$.

\begin{example}
\sloppy
In our toy example, $tokens(S_1)=tokens(r_1)=\{$\texttt{Cannon}, \texttt{EOS}, \texttt{1100D}, \texttt{Buy}, \texttt{Canon}, \texttt{32}, \texttt{mp}, \texttt{NP-400}, $\cdots\}$, while $tokens(S_2)$ is the union of $tokens(r_2)$ and $tokens(r_3)$.
\end{example}

\begin{table*}
    \centering
    \begin{tabular}{|c|l|}
    \hline 
         Symbol & Definition \\
    \hline\hline    
         $\mathcal{S}$ & a set of sources \\ \hline
         $S \in \mathcal{S}$ & a source \\ \hline
         $r \in S$ & a record \\ \hline
         $schema(r)$ & non-null attributes in a record $r \in S$ \\ \hline
         $schema(S)$ & non-null attributes in a source $S \in \mathcal{S}$ \\ \hline
         $tokens(r)$ & text tokens in a record $r \in S$ \\ \hline
         $tokens(S)$ & text tokens in a source $S \in \mathcal{S}$ \\ \hline
         $A$ &  non-null attributes in a set of sources $\mathcal{S}$ \\ \hline
         $T$ & target schema \\ \hline
         $V$ &  records in a set of sources $\mathcal{S}$ \\ \hline
         $M^+ \subseteq A \times T$ & matching attributes \\ \hline 
         $E^+ \subseteq V \times V$ & matching record pairs \\ \hline 
    \end{tabular}
    \caption{Notation table.\label{tab:notation}}
\end{table*}

Based on the presented data model, we now define the tasks and their variants considered in this work to illustrate the flexibility of our benchmark. %
For SM we consider two variants, namely: \emph{catalog-based} and \emph{mediated schema}. For ER we consider three variants, namely: \emph{self-join}, \emph{similarity-join} and a \emph{schema-agnostic} variant. Finally, we discuss the evaluation metrics used in our experiments. Table~\ref{tab:notation} summarizes the notation used in this section and in the rest of the paper.

\subsection{Schema Matching}  

Given a set of sources $\sources$, let $A = \bigcup_{S \in \sources} schema(S)$ and $T$ be a set of properties of interest. $T$ is referred to as the \emph{target schema} and can be either defined manually or set equal to the schema of a given source in $\sources$. Schema Matching is the problem of finding correspondences between elements of two schemata, such as $A$ and $T$~\cite{bernstein2011generic,rahm2001survey}. A formal definition is given below. 

\begin{definition}[Schema Matching] Let $M^+ \subseteq A \times T$ be a set s.t. $(a, t) \in M^+$ iff the attribute $a$ refers to the property $t$, SM can be defined as the problem of finding pairs in $M^+$.
\end{definition}

\begin{example}
\sloppy
Let $\sources=\{S_1,S_2\}$ and $A=schema(S_1) \cup schema(S_2)=\{a_1, \dots, a_{10}\}$ with 
$a_1=S_1.\texttt{battery}$, $a_2=S_1.\texttt{brand}$,
$a_3=S_1.\texttt{resolution}$, $a_4=S_1.\texttt{digital\_screen}$, 
$a_5=S_1.\texttt{size}$, 
$a_6=S_2.\texttt{battery\_model}$, $a_7=S_2.\texttt{battery\_chemistry}$, $a_8=S_2.\texttt{manufacturer}$, $a_9=S_2.\texttt{score}$ and $a_{10}=S_2.\texttt{size}$. 
Let $T=\{t_1, t_2, t_3\}$, with $t_1=$``battery model'', $t_2=$``battery chemistry'' and $t_3=$``brand''. Then $M^+=\{(a_1,t_1)$, $(a_1, t_2)$, $(a_2,t_3)$, $(a_6,t_1)$, $(a_7, t_2)$, $(a_8, t_3)\}$. 
\end{example}

\noindent We consider two popular variants for the SM problem. 

\begin{enumerate}
    \item \textbf{Catalog SM.} Given a set of sources $\sources = \{S_1, S_2, \dots\}$ and a \emph{Catalog} source $S^* \in \sources$ s.t. $T=schema(S^*)$, find the correspondences in $M^+$. It is worth mentioning that, in a popular instance of this variant, the Catalog source may correspond to a Knowledge Graph.
    \item \textbf{Mediated SM.} Given a set of sources $\sources = \{S_1, S_2, \dots\}$ and a \emph{mediated schema} $T$, defined manually with a selection of different real-world properties, find the correspondences in $M^+$.
\end{enumerate}

\noindent In both the variants above, $\sources$ can consist of one or several sources. State of the art algorithms (e.g., FlexMatcher~\cite{chen2018biggorilla}), typically consider one or two sources at a time, rather than the entire set of sources available.

\myparagraph{Challenges} The main challenges provided by our dataset for the SM problem are detailed below. 
\begin{itemize}
    \item \textbf{Synonyms.} Our dataset contains attributes with different names but referring to the same property, such as $S_1$.\texttt{brand} and $S_2$.\texttt{manufacturer};
    \item \textbf{Homonyms.} Our dataset contains attributes with the same names but referring to different properties, such as $S_1$.\texttt{size} (the size of the digital screen) and $S_2$.\texttt{size} (dimension of the camera body);
    \item \textbf{Granularity.} In addition to one-to-one correspondences (e.g., $S_1$.\texttt{brand} with $S_2$.\texttt{manufacturer}), our dataset contains one-to-many and many-to-many correspondences, due to different attribute granularities. $S_1$.\texttt{battery}, for instance, corresponds to both $S_2$.\texttt{battery\_model} and $S_2$.\texttt{battery\_chemistry}.
\end{itemize}

\subsection{Entity Resolution}

Given a set of sources $\sources$, let $V = \bigcup_{S \in \sources} S$ be the set of records in $\sources$. Entity Resolution (ER) is the problem of finding records referring to the same entity. Formal definition is given below. 

\begin{definition}[Entity Resolution] Let $E^+ \subseteq V \times V$ be a set s.t. $(u,v) \in E^+$ iff $u$ and $v$ refer to the same entity, ER can be defined as the problem of finding pairs in $E^+$. 
\end{definition}

We note that $E^+$ is transitively closed, i.e., if $(u,v) \in E^+$ and $(u,w) \in E^+$, then $(v,w) \in E^+$, and each connected component $\{C_1, \dots, C_k\}$ is a clique representing a distinct entity. We call each clique $C_i$ a \textit{cluster} of $V$. %

\begin{example}
Let $\sources=\{S_1,S_2\}$ and $V=\{r_1, r_2, r_3\}$. Then $E^+=\{(r_1,r_2)\}$ and there are two clusters, namely $C_1=\{r_1,r_2\}$ representing a ``Canon EOS 1100D'' and $C_2=\{r_3\}$ representing a ``Sony A7''.
\end{example}

\noindent We consider three variants for the ER problem.

\begin{enumerate}
    \item 
    \textbf{Similarity-join ER.} Given two sources $\sources = \{S_{left}, S_{right}\}$ find the record pairs in $E^+_{sim} \subseteq E^+$,  $E^+_{sim} \subseteq S_{left} \times S_{right}$.
    \item 
    \textbf{Self-join ER.} Given a set of sources $\sources = \{S_{1}, S_2 \dots\}$ find the record pairs in $E^+$.
    \item 
    \textbf{Schema-agnostic ER.} In both the variants above, we assume that a solution for the SM problem is available, that is, attributes of every input record for ER are previously aligned to a manually-specified mediated schema $T$. This variant is the same as self-join ER but has no SM information available.
\end{enumerate}

\myparagraph{Challenges} The main challenges provided by our dataset for the ER problem are detailed below.

\begin{itemize}
    \item \textbf{Variety.} Different sources may use different format and naming conventions. For instance, the records $r_1$ and $r_2$ conceptually have the same value for the \texttt{resolution} attribute, but use different formats. In addition, records can contain textual descriptions such as the battery in $r_1$ and data sources of different countries can use different conventions (i.e., inches/cm and ``EOS''/``Rebel''). As a result, records referring to different entities can have more similar representation than records referring to the same entity. %
    \item \textbf{Noise.} Records can contain noisy or erroneous values, due to entity misrepresentation in the original web page and in the data extraction process. An example of noise in the web page is ``Cannon'' in place of ``Canon'' in the page title of $r_1$. An example of noise due to data extraction is the rating in $r_3$, which was extracted from a different portion of the web page than the product's specification.
    \item \textbf{Skew.} Finally, the cluster size distribution over the entire set of sources is skewed, meaning that some entities are over-represented and others are under-represented. %
\end{itemize}

\subsection{Performance measures}

We use the standard performance measures of precision, recall and F1-measure for both the SM and ER tasks. More specifically, let $M^- = (A \times T) \setminus M^+$ and $E^- = (V \times V) \setminus E^+$. For each of the tasks defined in this paper, we consider manually verified subsets of $M^+$, $M^-$, $E^+$ and $E^-$ that we refer to as \emph{ground truth}, and then we define precision, recall and F1-measure with respect to the manually verified ground truth. 

Let such ground truth subsets be denoted as $M^+_{gt}$, $M^-_{gt}$, $E^+_{gt}$ and $E^-_{gt}$. In our benchmark we ensure that $M^+_{gt} \cup M^-_{gt}$ yields a complete bipartite sub-graph of $A \times T$ and $E^+_{gt} \cup E^-_{gt}$ yields a complete sub-graph of $V \times V$. More details on the ground truth can be found in Section~\ref{sec:profiling}.

Finally, in case the SM or ER problem variant at hand has prior restrictions (e.g., a SM restricted to one-to-one correspondences or a similarity-join ER restricted to inter-source pairs), we only consider the portion of the ground truth satisfying those restrictions.

\section{Profiling}
\label{sec:profiling}

\begin{table*}[th]
\centering
\begin{tabular}{|l|c|c|c|c|c|c|c|c|c|}
\hline
 \multirow{2}{*}{vertical $v$} & \multirow{2}{*}{$|\mathcal{S}_{v}|$} & \multirow{2}{*}{$|V|$} & \multirow{2}{*}{$|S_{max}|$} & \multirow{2}{*}{$|A|$} & \multirow{2}{*}{$a_{avg}$} & \multirow{2}{*}{$t_{avg}$} & \multicolumn{3}{c|}{ground truth} \\ 
\cline{8-10}
 & & & & & & & $|T^*|$ & $k^*$ & $|C^*_{max}|$  \\
\hline\hline
\texttt{camera} & 24 & 29,787 & 14,274 & 4,660 & 16.74 & 8.53 & 56 & 103 & 184 \\ \hline
\texttt{monitor} & 26 & 16,662 & 4,282 & 1,687 & 25.64 & 4.05 & 87 & 232 & 33 \\ \hline
\texttt{notebook} & 27 & 23,167 & 8,257 & 3,099 & 29.42 & 9.85 & 44 & 208 & 91 \\ \hline
\end{tabular}
\caption{Verticals in the \alaska{} dataset.}
\label{tab:profiling_table}
\end{table*}

\alaska{} contains data from three different e-commerce domains: \texttt{camera}, \texttt{monitor}, and \texttt{notebook}. We refer to such domains as \emph{verticals}. Table~\ref{tab:profiling_table} reports, for each vertical $v$, the number of sources $|\mathcal{S}_{v}|$, the number of records $|V|$, the number of records in the largest source $|S_{max}|$, $S_{max}=\textnormal{argmax}_{S \in \mathcal{S}_{v}} |S|$, the total number of attributes $|A|$, and the average number of record attributes in each vertical $a_{avg}=\textnormal{avg}_{r \in V} |schema(r)|$, the average number of record tokens $t_{avg}=\textnormal{avg}_{r \in V} |tokens(r)|$ (i.e., the record length). 
Table~\ref{tab:profiling_table} also shows details about our manually curated ground truth: the number of attributes $|T^*|$ in the target schema for the SM task; the number of entities $k^*$ and the size of the largest cluster $|C^*_{max}|$ for the ER task. Note that for each source we provide the complete set of attributes and record matches with respect to the mentioned target schema and entities.

In the rest of this section, we provide profiling metrics for our benchmark and show their distribution over \alaska{} sources. Profiling metrics include: $(i)$~traditional size metrics such as the number of records and attributes, and $(ii)$~three new metrics, dubbed \emph{Attribute Sparsity}, \emph{Source Similarity} and \emph{Vocabulary Size}, that quantify different dimensions of heterogeneity for one or several sources. The goal of the considered profiling metrics is to allow users of our benchmark to pick and choose different subsets of sources for each vertical so as to match their desired use case. Possible scenarios with different difficulty levels include for instance integration tasks over head sources, tail sources, sparse sources, dense sources, clean sources, dirty sources and so on. %

\begin{figure}
    \centering
    \begin{subfigure}{0.85\columnwidth}
    \includegraphics[width=\textwidth]{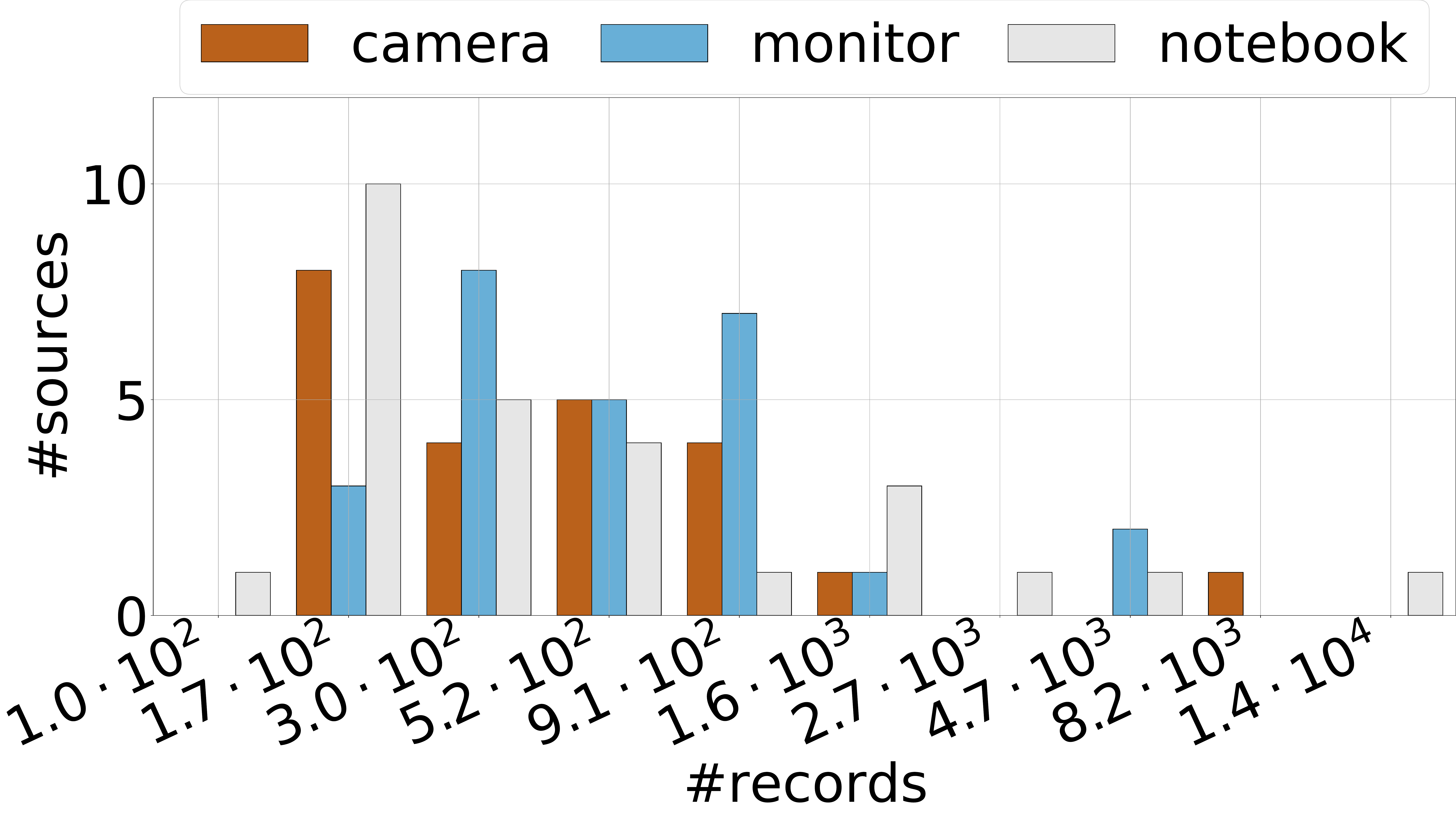}
    \caption{\label{fig:records}}
    \end{subfigure}

    \begin{subfigure}{0.85\columnwidth}
    \includegraphics[width=\textwidth]{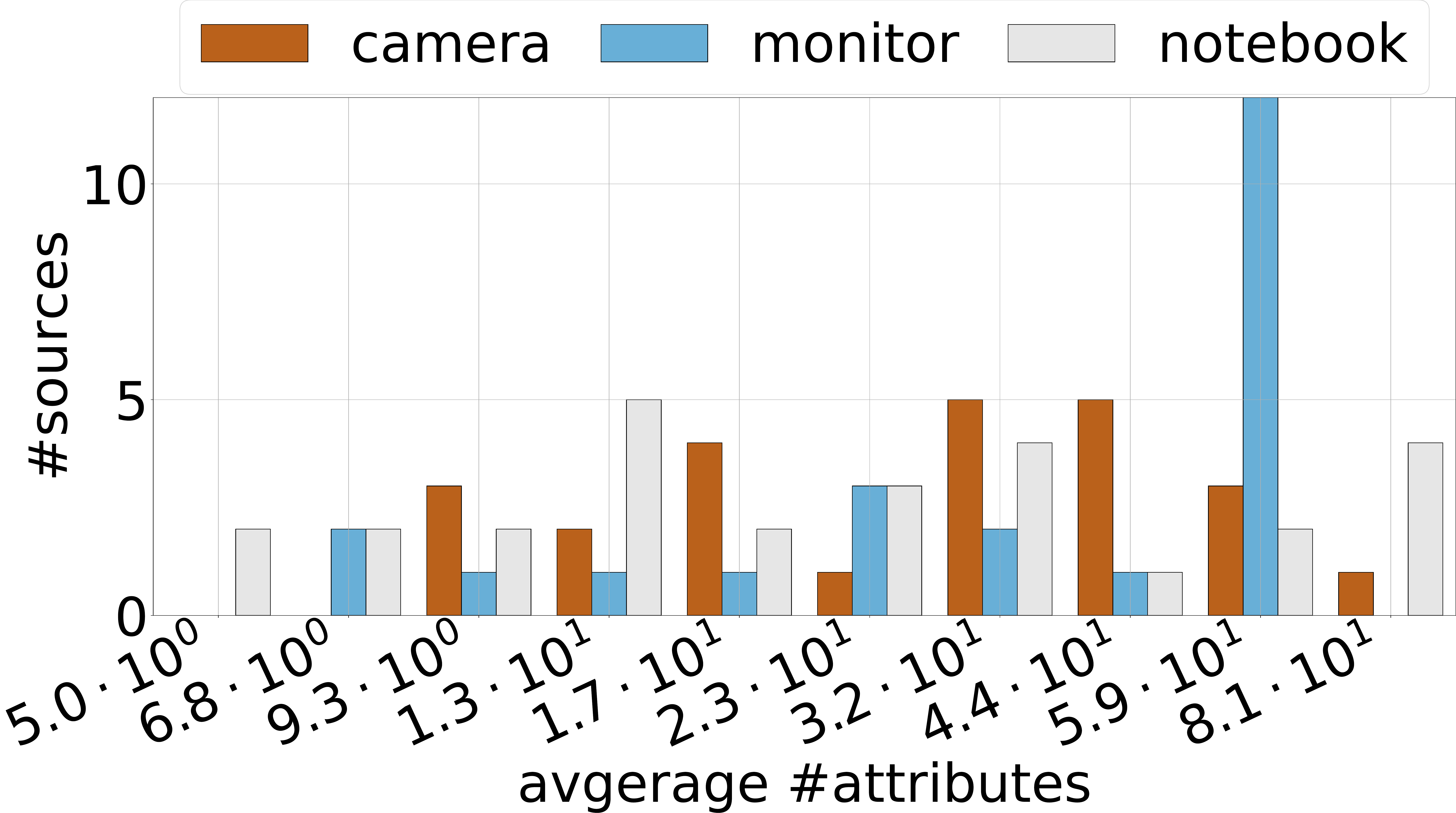}
    \caption{\label{fig:attributes}}
    \end{subfigure}
    \caption{(a) Number of records and (b) average number of attributes per source. Note that the $x$-axis is in log scale, where the value range on the $x$-axis is divided into 10 buckets and each bucket is a constant multiple of the previous bucket.} %
    \label{fig:volume}
\end{figure}

\myparagraph{Size}  Figures~\ref{fig:records} and~\ref{fig:attributes} plot the distribution of \alaska{} sources with respect to the number of records ($|S|$) and the average number of attributes per record %
($a_{avg}=\textnormal{avg}_{r \in S} |schema(r)|$), respectively. Figure~\ref{fig:records} shows that the \texttt{camera} and \texttt{notebook} verticals have a few big sources and a long tail of small sources, while the \texttt{monitor} vertical have more medium-sized sources. As for the number of attributes, Figure~\ref{fig:attributes} shows that \texttt{camera} and \texttt{notebook} follow the same distribution, while \texttt{monitor} has notably more sources with higher average number of attributes.

\begin{figure}
    \centering
    \begin{subfigure}{0.8\columnwidth}
    \includegraphics[width=\columnwidth]{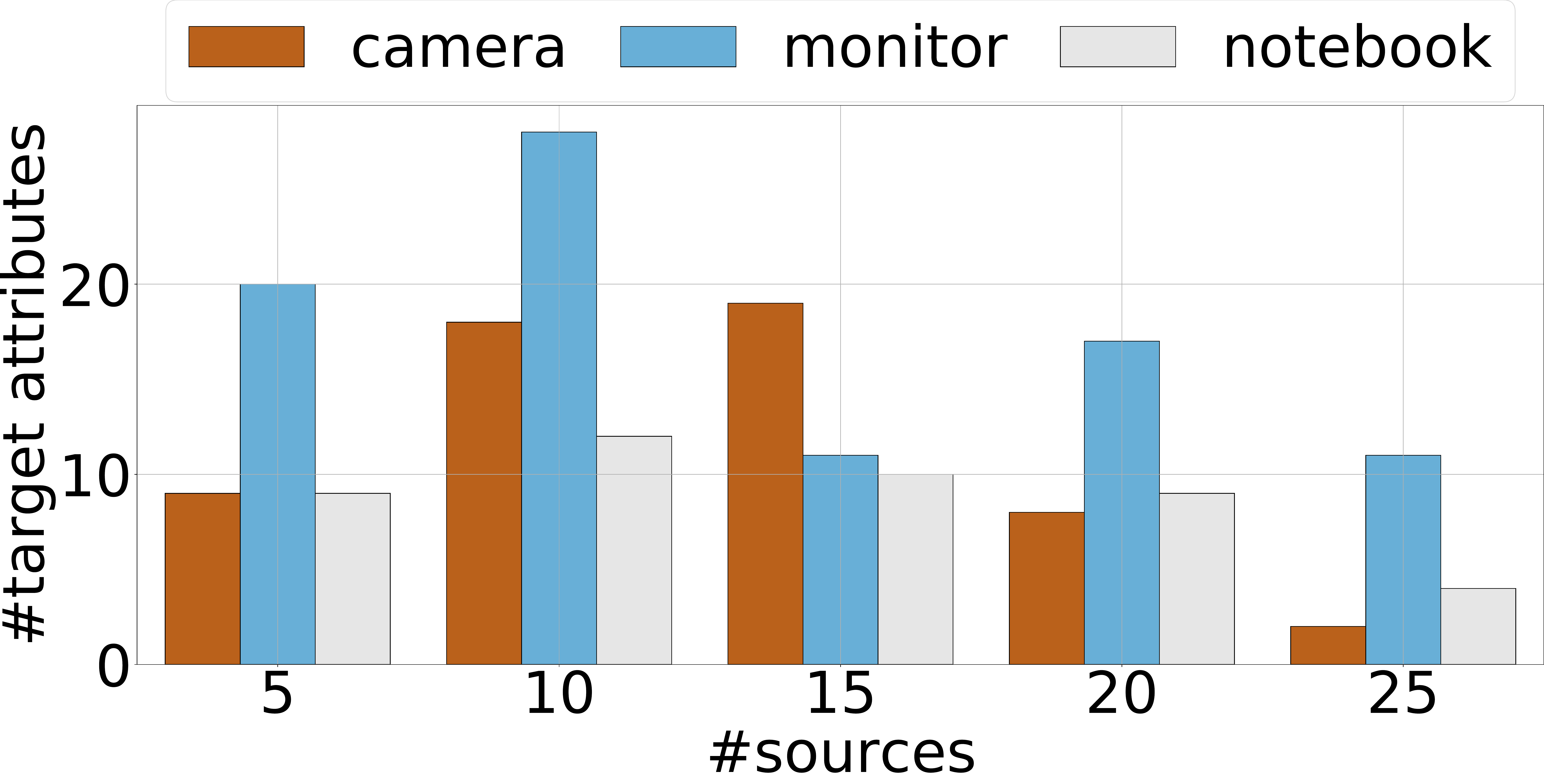}
    \caption{\label{fig:target-attributes-frequency}}
    \end{subfigure}
    
    \begin{subfigure}{0.8\columnwidth}
    \includegraphics[width=\columnwidth]{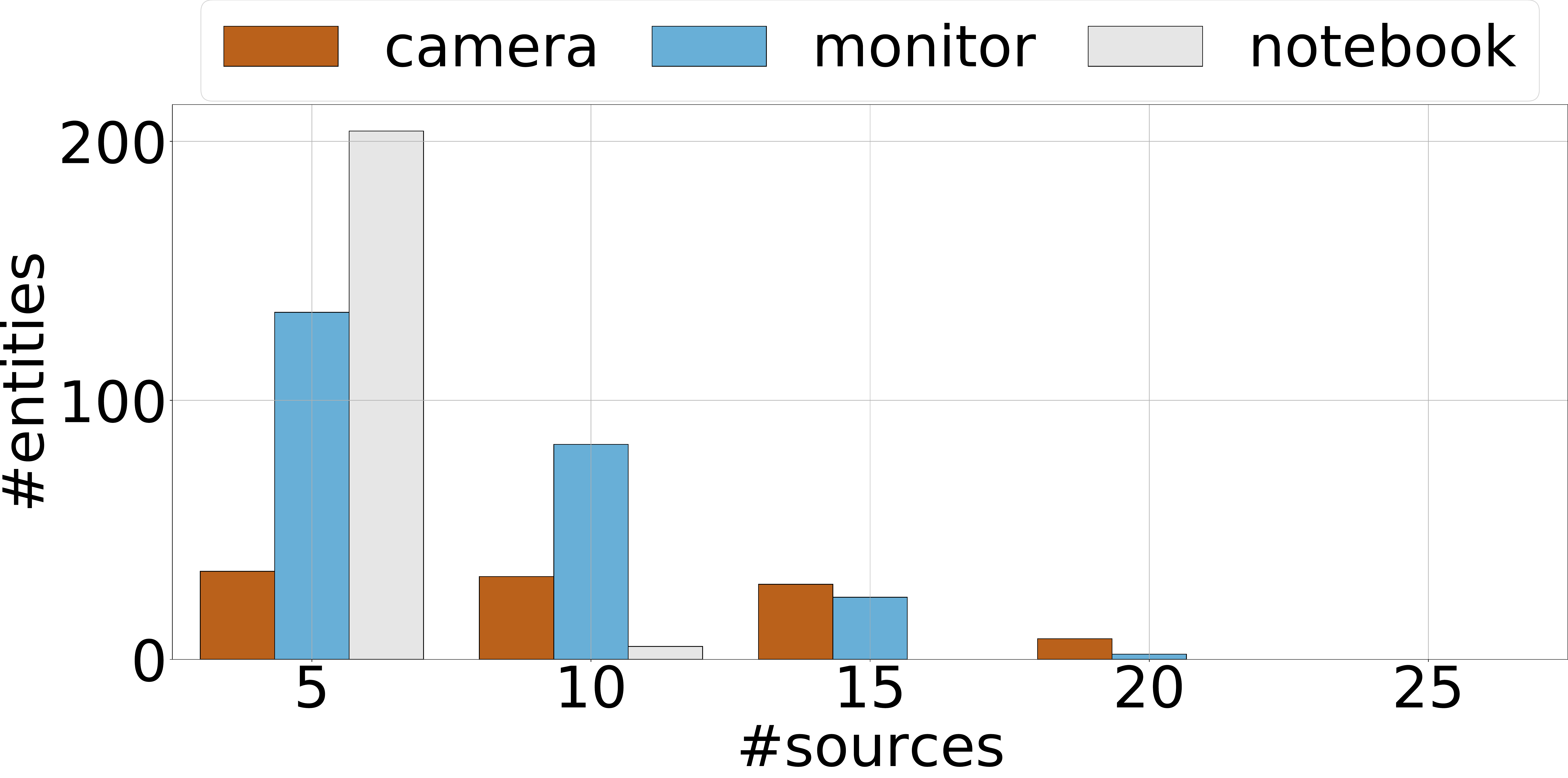}
    \caption{\label{fig:entity-attributes-frequency}}
    \end{subfigure}
    
    \caption{(a) Target attributes and (b) entities by number of \alaska{} sources in which they are mentioned. \label{fig:gtfrequency}}
\end{figure}

\begin{figure}
    \centering
    \includegraphics[width=0.9\columnwidth]{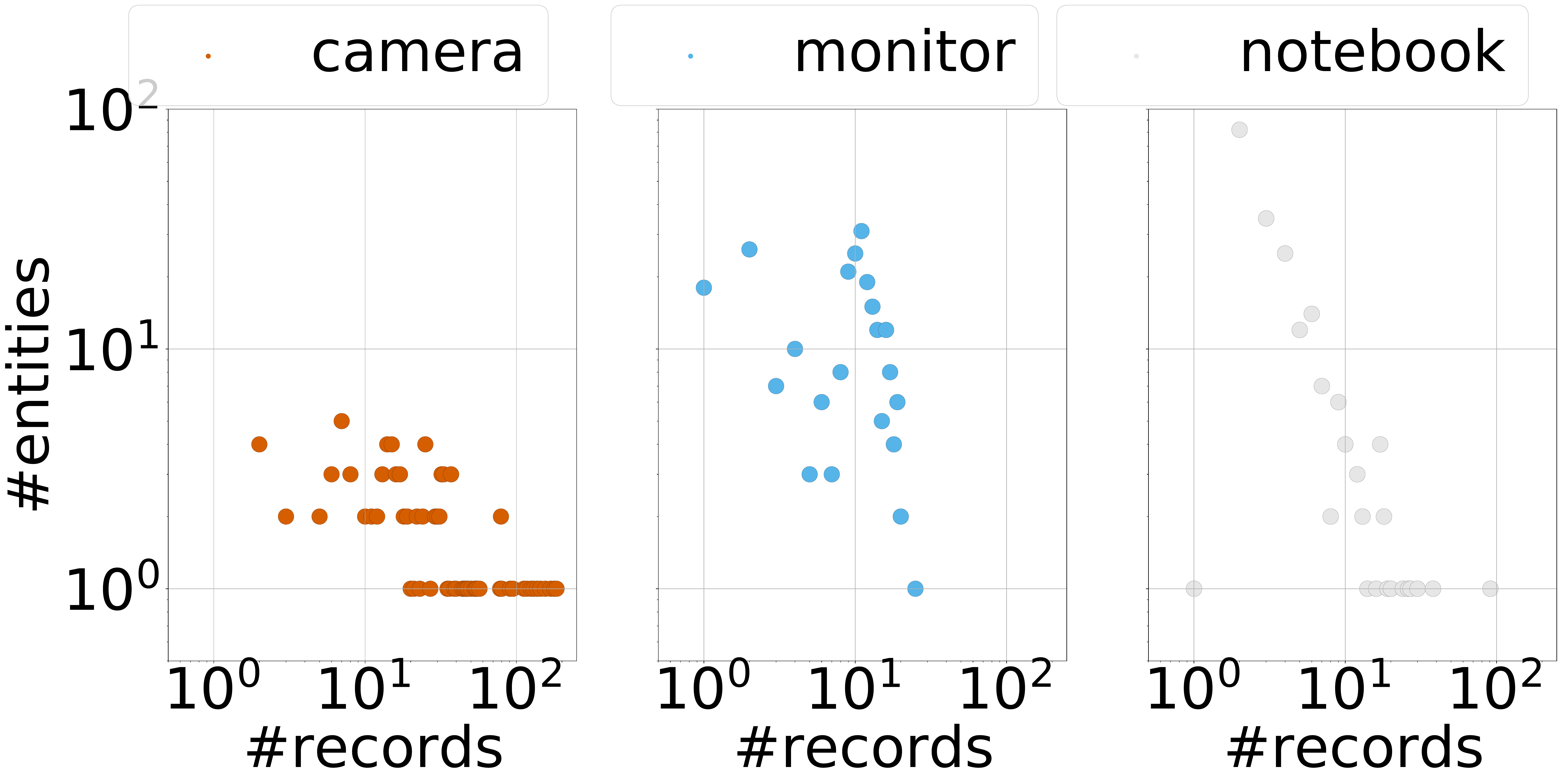}
   \caption{Cluster size distribution.}
    \label{fig:entities-frequency}
\end{figure}

Figure~\ref{fig:gtfrequency} reports the distribution of target attributes (\ref{fig:gtfrequency}a) and entities (\ref{fig:gtfrequency}b) in our manually curated ground truth, with respect to the number of sources in which they appear. It is worth noticing that a significant fraction of attributes in our ground truth are present in most sources, but there are also tail attributes that are present in just a dfew sources (less than $10\%$). Regarding entities, observe that in \texttt{notebook} most entities span less than $20\%$ of sources, while in \texttt{monitor} and \texttt{camera} there are popular entities that are present in up to $80\%$ of sources.
Figure~\ref{fig:entities-frequency} shows the cluster size distribution of entities in the ground truth. Observe that the three \alaska{} verticals are significantly different: \texttt{camera} has the largest cluster (184 records) %
\texttt{monitor} has both small and medium-size clusters, and \texttt{notebook} has the most skewed cluster size distribution, with many small clusters. %

\begin{figure}
    \centering
    \includegraphics[width=0.85\columnwidth]{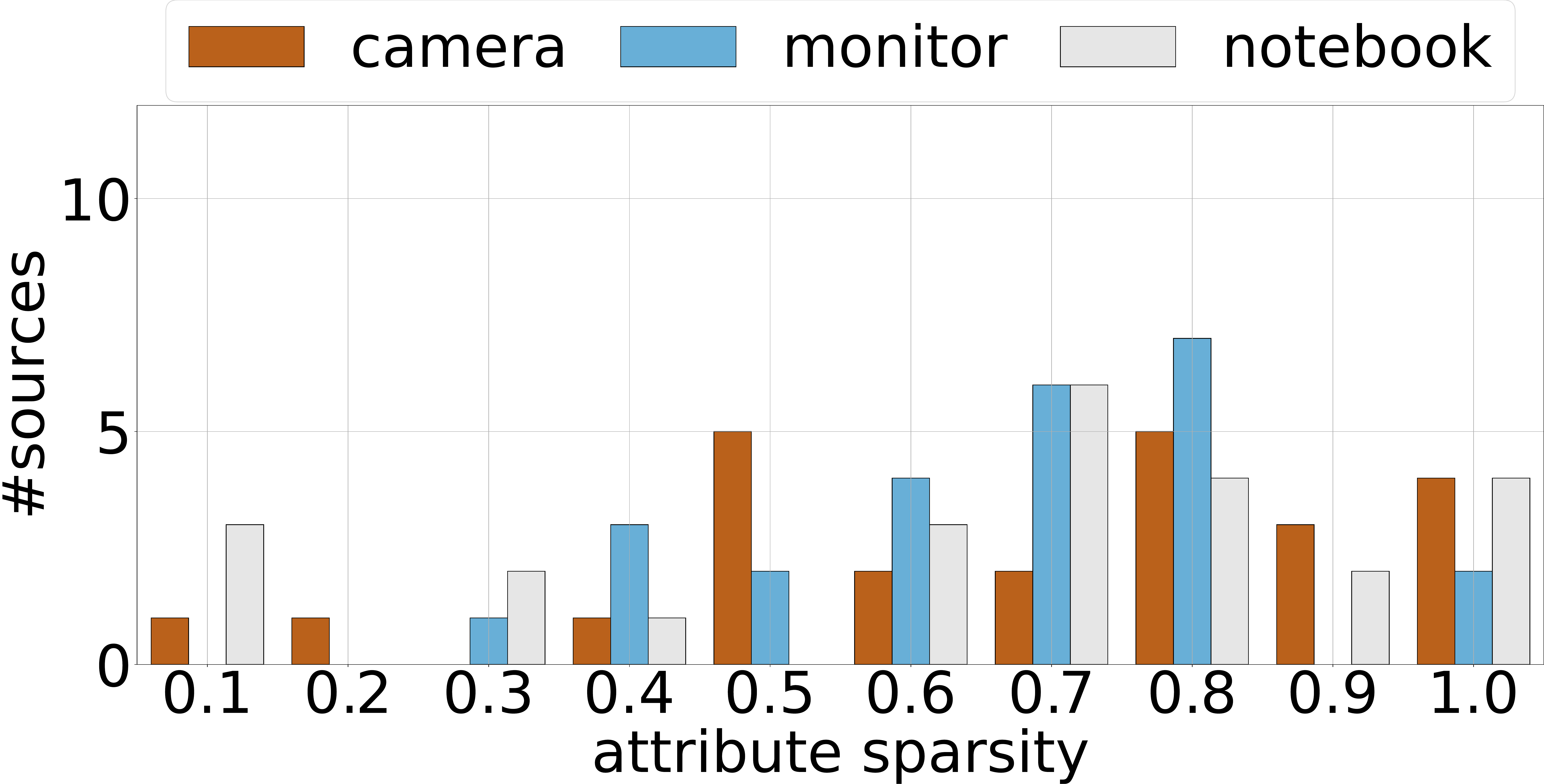}
    \caption{Source attribute sparsity, from dense to sparse.}
    \label{fig:sparsity}
\end{figure}

\myparagraph{Attribute Sparsity} Attribute sparsity aims at measuring how often the attributes of a source are actually used by its records. Attribute Sparsity, $AS: \mathcal{S} \rightarrow [0, 1]$, is defined as follows:
\begin{equation}
  AS(S) = 1 - \frac{\sum_{r \in S} |schema(r)|}{|schema(S)| \cdot |S|} 
  \label{eq:as}
\end{equation}

\noindent 
In a \emph{dense} source (low $AS$), most records have the same set of attributes. In contrast, in a \emph{sparse} source (high $AS$), most attributes are null for most of the records. Denser sources represent easier instances for both SM and ER, as they provide a lot of non-null values for each attribute and more complete information for each record.

Figure~\ref{fig:sparsity} shows the distribution of \alaska{} sources according to their AS value: \texttt{notebook} has the largest proportion of dense sources ($AS < 0.5$), while \texttt{monitor} has the largest proportion of sparse sources ($AS>0.5$).

\begin{figure}
    \centering
    \includegraphics[width=0.85\columnwidth]{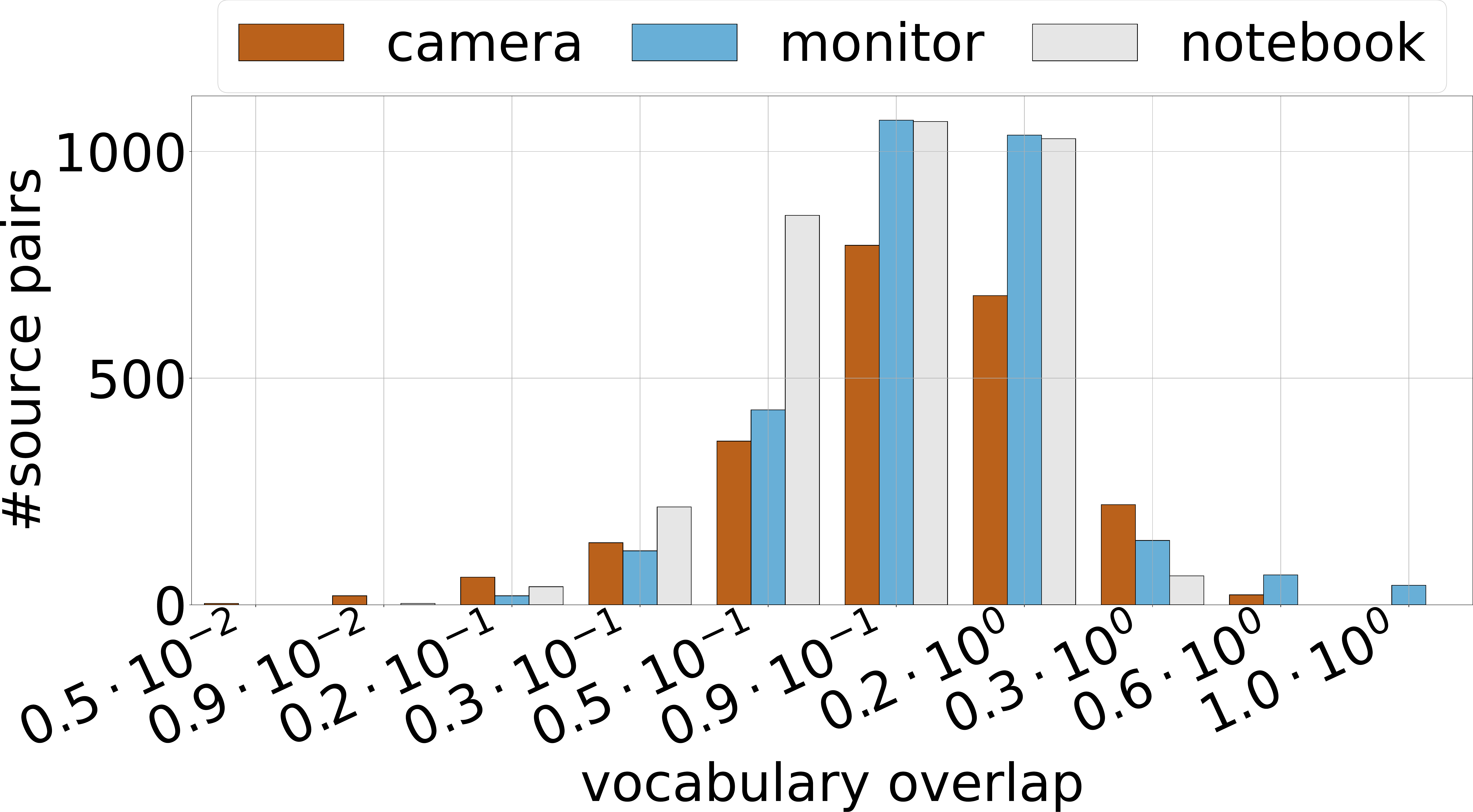}
    \caption{Pair-wise Source Similarity, from less similar to more similar pairs. Note that the $x$-axis is in log scale, namely, each bucket is a constant multiple of the previous bucket.}
    \label{fig:similarity}
\end{figure}

\myparagraph{Source Similarity} Source Similarity has the objective of measuring the similarity of two sources in terms of attribute values. Source similarity, $SS: \mathcal{S} \times \mathcal{S} \rightarrow [0, 1]$, is defined as the Jaccard index of the two source vocabularies, as follows: 
\begin{equation}
    SS(S_1,S_2) = \frac{|tokens(S_1) \cap tokens(S_2)|}{|tokens(S_1) \cup tokens(S_2)|}
\end{equation}

\noindent Source sets with high pair-wise SS values have typically similar representations of entities and properties, and are easy instances for SM.

Figure~\ref{fig:similarity} shows the distribution of source pairs according to their SS value. As for SS, in all three \alaska{} verticals the majority of source pairs have medium-small overlap. In \texttt{camera} there are notable cases with SS close to zero, providing extremely challenging instances for SM. Finally, in \texttt{notebook} there are pairs of sources that use approximately the same vocabulary ($SS$ close to one) which can be used as a frame of comparison when testing vocabulary-sensitive approaches.

\begin{figure}
    \centering
    \includegraphics[width=0.85\columnwidth]{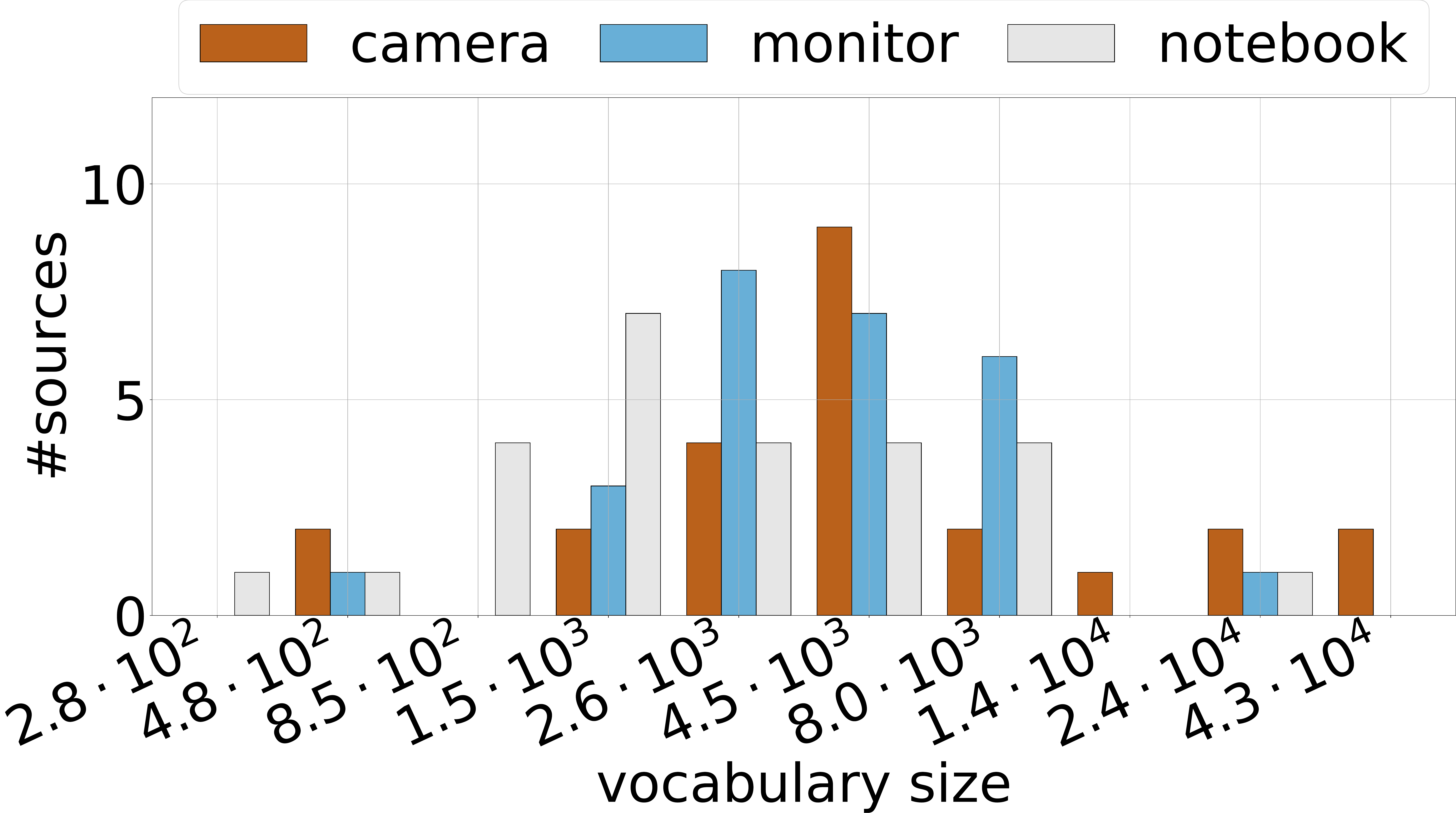}
    \caption{Source vocabulary size, from concise to verbose. Note that the $x$-axis is in log scale. 
    \label{fig:verb}}
\end{figure}

\myparagraph{Vocabulary Size} Vocabulary Size, $VS: S \rightarrow \mathbb{N}$, measures the number of distinct tokens that are present in a source:
\begin{equation}
    VS(S) = |tokens(S)|
\end{equation}

\noindent Sources can have high VS for multiple reasons, including values with long textual descriptions, or attributes with large ranges of categorical or numerical values. In addition, large sources with low sparsity typically have higher VS than small ones. Sources with high $VS$ usually represent easier instances for the schema-agnostic ER task.

Figure~\ref{fig:verb} shows the distribution of \alaska{} sources according to their VS value. The \texttt{camera} vertical has the most verbose sources, while \texttt{notebook} has the largest proportion of sources with smaller vocabulary. Therefore, \texttt{camera} sources are more suitable on average for testing NLP-based approaches for ER such as~\cite{mudgal2018deep}. %

\section{Experiments}
\label{sec:casestud}

We now describe a collection of significant usage scenarios for our \alaska{} benchmark. In each scenario, we consider one of the tasks in Section~\ref{sec:prel} and show the performance of a representative method for that task over different source selections, with different profiling properties.

Our experiments were performed on a server machine equipped with a CPU Intel Xeon E5-2699 chip with 22 cores running 44 threads at speed up to 3.6 GHz, 512 GB of DDR4 RAM and 4 GPU NVIDIA Tesla P100-SXM2 each with 16 GB of available VRAM  space. The operating system was Ubuntu 17.10, kernel version 4.13.0, Python version 3.8.1 (64-Bit) and Java 8.

\subsection{Representative methods}

Let us illustrate the representative methods that we have considered for demonstrating the usage scenarios for our benchmark, and the settings used in our experiments.  %

\myparagraph{Schema Matching Methods}  We selected two methods for the Schema Matching tasks, namely {\sacatalog}~\cite{nguyen2011synthesizing} and {\flex}~\cite{chen2018biggorilla}.
\begin{itemize}
    \item {\sacatalog}~\cite{nguyen2011synthesizing} is a system designed to enrich product catalogs with external specifications retrieved from the Web. This system addresses several issues and includes a catalog SM algorithm. We re-implemented this SM algorithm and used it in our catalog SM experiments. The SM algorithm in {\sacatalog} computes a similarity score between the attributes of an external source and the attributes of a catalog; then the algorithm matches attribute pairs with the highest score. The similarity score is computed by means of a classifier, which is trained with pairs of attributes with the same name, based on the assumption that they represent positive correspondences. It is important to observe that some of these pairs could actually be false positives due to the presence of homonyms. In our experiment, each run of the {\sacatalog} algorithm required less than 10 minutes.
    \item {\flex}~\cite{chen2018biggorilla} is one of the tools included in BigGorilla,\footnote{\url{https://biggorilla.org}} a broad project that gathers solutions for data integration and data preparation. We use the original Python implementation\footnote{\url{https://pypi.org/project/flexmatcher} version 1.0.4} in our mediated SM experiments. As suggested by the {\flex} documentation,\footnote{\url{https://flexmatcher.readthedocs.io}} we use two sources for training and a source for test, and kept the default configuration. %
    {\flex} uses the training sources to train multiple classifiers by considering several features including attribute names and attribute values. %
    A meta-classifier eventually combines the predictions made by each base classifier
    to make a final prediction. It detects the most important similarity features for each attribute of the mediated schema. In our experiments, the training step required from 2 to 5 minutes, and the test step from 10 seconds to 25 minutes, depending on the size of input sources. 
\end{itemize}    
    
\myparagraph{Entity Resolution Methods}  
In principle, our benchmark can be used for evaluating a variety of ER techniques. We have selected two recent (and popular) Deep Learning solutions, that have been proved very effective, especially when dealing with textual data, namely, DeepMatcher~\cite{mudgal2018deep} and DeepER~\cite{ebraheem2017deeper}.
\begin{itemize}
    \item DeepMatcher~\cite{mudgal2018deep} is a suite of Deep Learning models specifically tailored for ER. We used  %
    the Hybrid model of DeepMatcher for our self-join ER and similarity-join ER experiments. This model, which performed best in the original paper, uses a bidirectional RNN with decomposable attention. We used the original DeepMatcher Python implementation.\footnote{\url{https://github.com/anhaidgroup/deepmatcher}} 

    \item DeepER~\cite{ebraheem2017deeper} is a schema-agnostic ER system based on a distributed representation of tuples. We re-implemented the DeepER version with RNN and LSTM, which performed best in the original paper, and used it for our schema-agnostic ER experiments. %
   
\end{itemize}

We trained and tested both DeepMatcher and DeepER on the same selection of the data. For each experiment, we split the ground truth into training (60\%), validation (15\%) and test (25\%).  The results, expressed through the classic precision, recall and F1-measure metrics, are calculated on the test set. Since our ER ground truths are transitively closed, so are the training, validation and test sets. For the self-join ER tasks (schema-based and schema-agnostic) we down-sample the non-matching pairs in the training set to reduce imbalance between the size of matching and that of non-matching training samples. In particular, for each matching pair $(u,v)$, we sampled two non-matching pairs $(u,v')$ and ($u',v)$ with high similarity (i.e., high Jaccard index of token sets). %
DeepMatcher training required 30 minutes on average on the similarity-join ER tasks, and 1 hour on average on the self-join ER tasks. DeepER training required 25 minutes on average on the schema-agnostic ER tasks.

\subsection{Usage scenarios}

\begin{table*}
\centering
\small
\begin{tabular}{|l|c|c|p{7cm}|}
\hline
Task                & Method & Vertical &  Main source selection criteria \\ \hline\hline
Catalog SM          & \sacatalog{} & \texttt{monitor} & catalog sources with decreasing number of records \\ \hline
Mediated SM         & \flex{}   & \texttt{camera}   &  sources with different attribute sparsity and source similarity \\ \hline
Similarity-join ER  & DeepMatcher & \texttt{camera}    & sources with increasing attribute sparsity                 \\ \hline
Self-join ER        & DeepMatcher & \texttt{notebook} &  sources with increasing attribute sparsity and skewed cluster size distribution                 \\ \hline
Schema-agnostic ER  & DeepER &  \texttt{notebook} & sources with increasing vocabulary size and skewed cluster size distribution  \\ \hline
\end{tabular}
\caption{Usage scenarios and main source selection criteria. \label{tab:usage}}
\end{table*}

In the following we show the results of our representative selection of state-of-the-art methods over different usage scenarios. In each scenario, we consider one of the ER and SM variants described in Section~\ref{sec:prel} and show the performance of a method for the same task over three (or more) different source selections. Sources are selected according to the profiling metrics introduced in Section~\ref{sec:profiling} so as to yield task instances with different levels of difficulty (e.g., from easy to hard). For the sake of completeness, we use different verticals of the benchmark for different experiments. Results over other verticals are analogous and therefore are omitted. Usage scenarios discussed in this section are summarized in Table~\ref{tab:usage}.\footnote{For the tasks \emph{similarity-join ER} and \emph{self-join ER} we use the SM ground truth.}

\begin{figure}
    \centering
    \begin{subfigure}{0.8\columnwidth}
    \includegraphics[width=\textwidth]{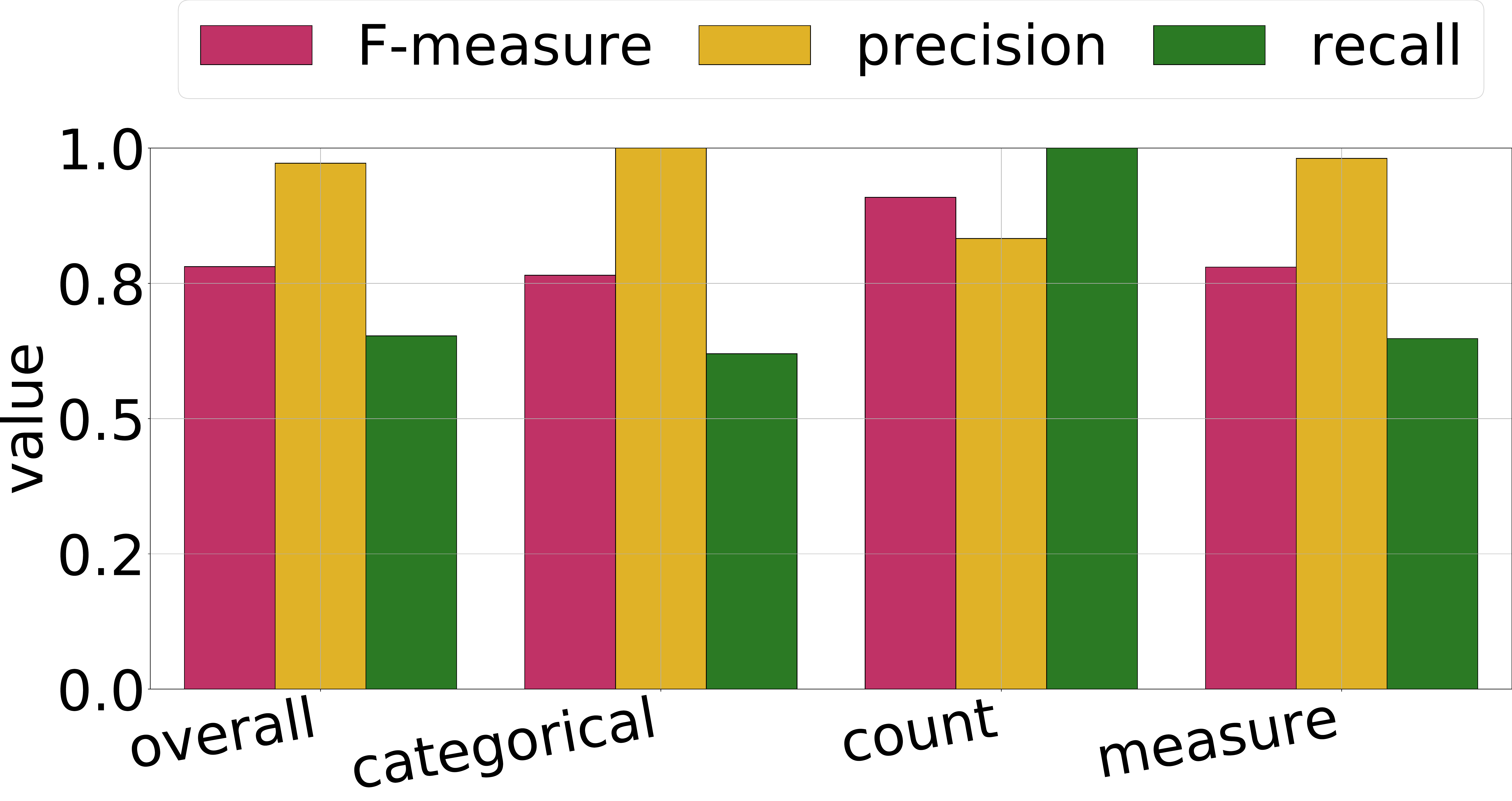}
    \caption{$S^*$ = jrlinton}
    \end{subfigure}
    
    \begin{subfigure}{0.8\columnwidth}
    \includegraphics[width=\textwidth]{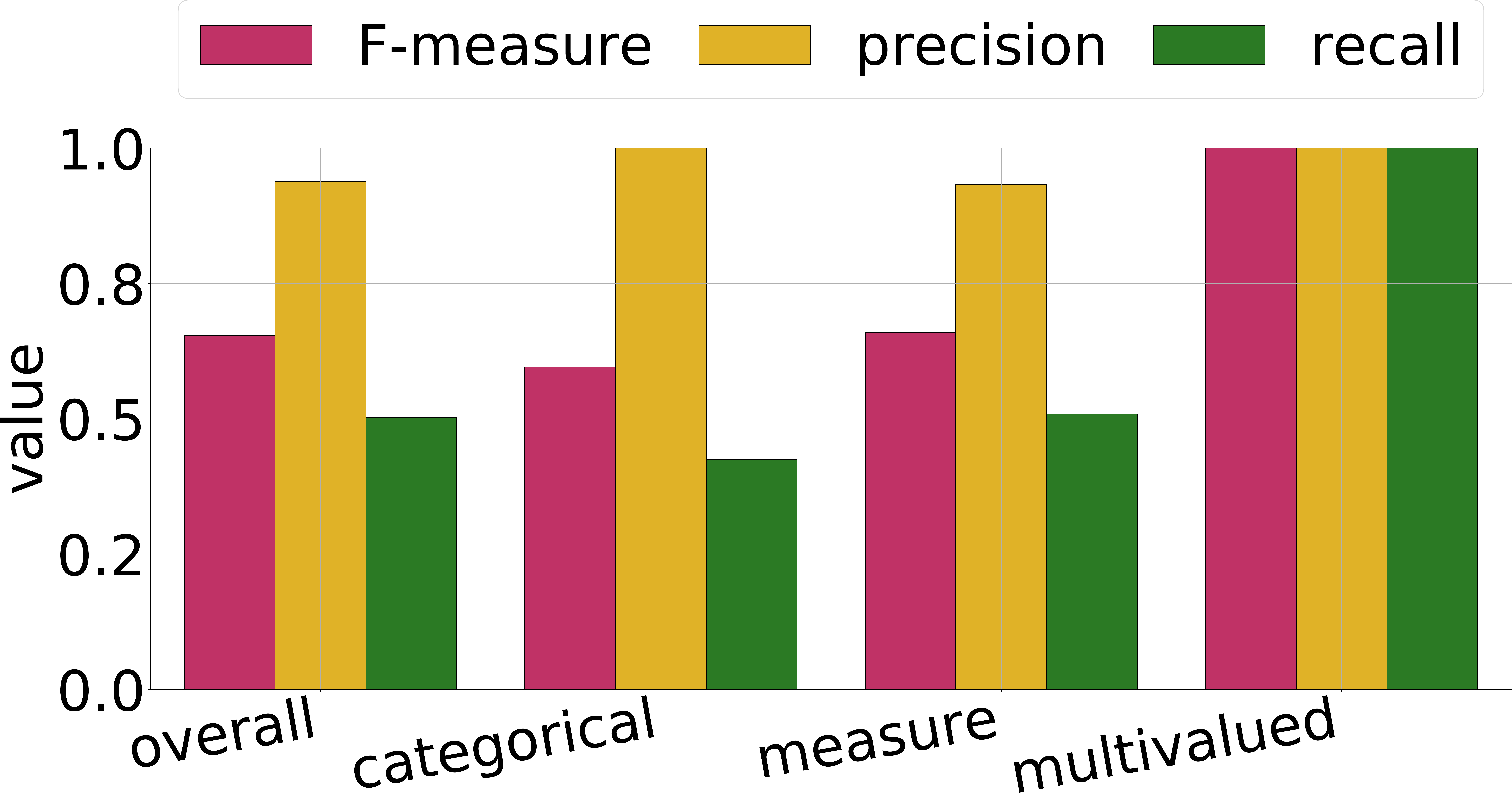}
    \caption{$S^*$ = vology}
    \end{subfigure}
    
    \begin{subfigure}{0.8\columnwidth}
    \includegraphics[width=\textwidth]{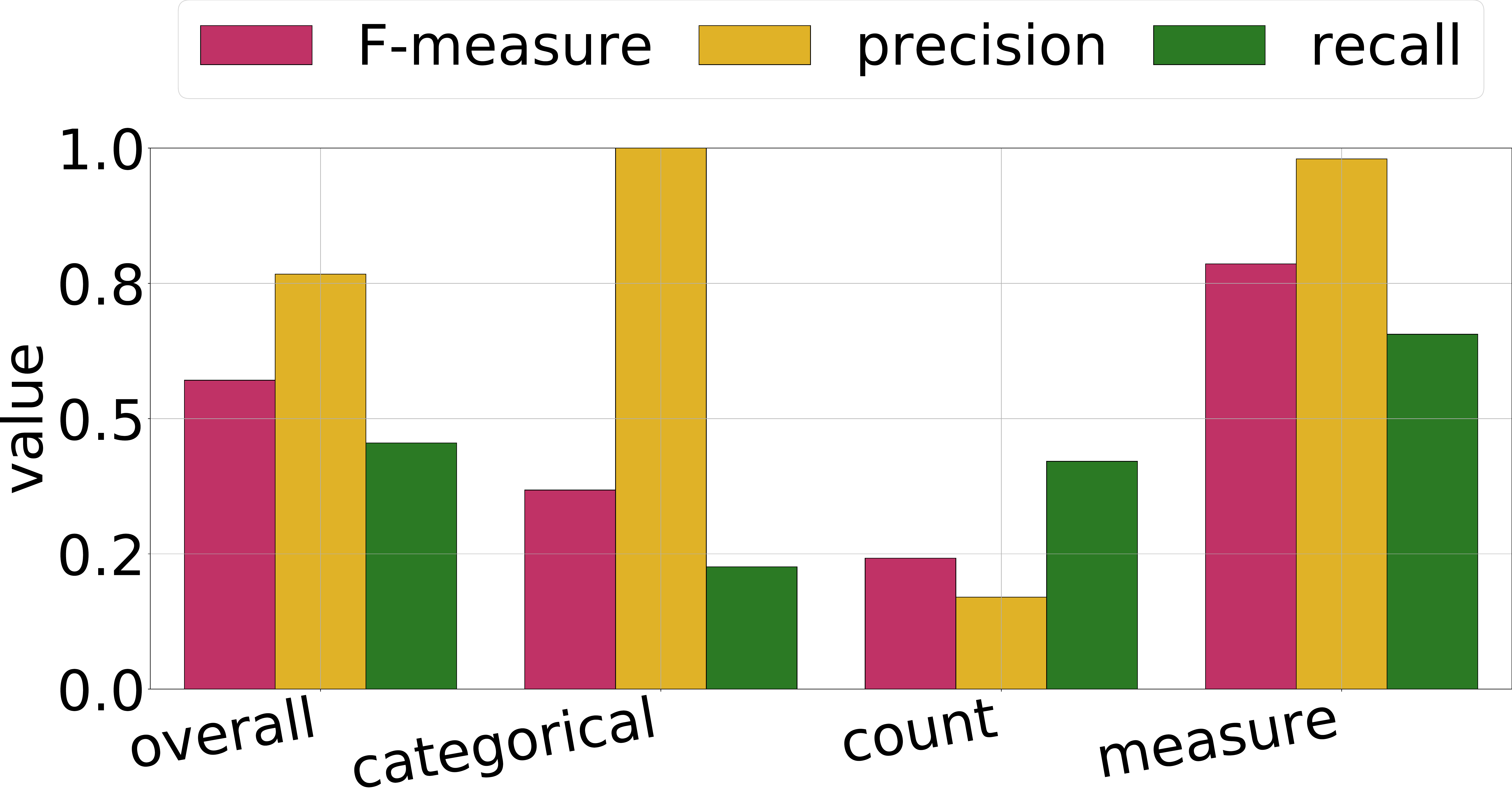}
    \caption{$S^*$ = nexus-t-co-uk}
    \end{subfigure}    
    \caption{\sacatalog{} results for the catalog SM task on different instances from the \texttt{monitor} vertical.}
    \label{fig:agrawal}
\end{figure}

\subsubsection{Schema Matching Scenarios}
\noindent
\textbf{Catalog SM.} Catalog SM approaches are conceived to enrich a target clean source, referred to as \emph{catalog}, with many other, possibly dirty, sources. Larger catalogs (i.e., with a larger number of records) typically yield better performance, as more redundancy can be exploited. 
For this task, we consider the \sacatalog{}  algorithm~\cite{nguyen2011synthesizing}. 

Figure~\ref{fig:agrawal} shows \sacatalog{} performance over three different choices of the catalog source $S^*$ in the \texttt{monitor} vertical. We select catalog sources with decreasing size and reasonably low attribute sparsity ($AS(S^*) \leq 0.5$), as reported in Table~\ref{tab:nguyen}. For each run, we set $\mathcal{S}=\mathcal{S}_{monitor} \setminus S^*$. 

\begin{table}
\centering
\begin{tabular}{|l|c|c|c|}
\hline
 $S^*$ & $|S^*|$ & $AS(S^*)$ & Difficulty \\ \hline \hline
jrlinton & 1,089 & 0.50 & easy \\ \hline
vology & 528 & 0.46 & medium \\ \hline
nexus-t-co-uk & 136 & 0.30 & hard \\ \hline
\end{tabular}
\caption{Sources considered for our catalog SM experiments.}
\label{tab:nguyen}
\end{table}

As expected, the overall F-measure is lower for smaller catalog sources. We also observe that even in the same catalog, some attributes can be much easier to match than others. %
For this reason, in addition to the overall F-measure, we also report the partial F-measure of \sacatalog{} when considering only attributes of the following types.
\begin{itemize}
    \item \textbf{Measure} indicates an attribute providing a numerical value (like cm, inch...). 
    \item \textbf{Count} indicates a quantity (number of USB ports...).
    \item \textbf{Single-value} indicates categorical attributes providing a single value for a given property (e.g., the color of a product). 
    \item \textbf{Multivalued} are categorical attributes with multiple possible values (e.g., the set of ports of a laptop).
\end{itemize}
Finally, we note that thanks to the redundancy in larger catalogs, \sacatalog{} can be robust to moderate and even to high attribute sparsity. However, when sparsity become too high, performances can quickly decrease. For instance, we run PSE using eBay as catalog source (99\% sparsity), and reported 0.33 as F-measure, despite the large-size source.

\begin{figure}
    \centering
    \includegraphics[width=0.85\columnwidth]{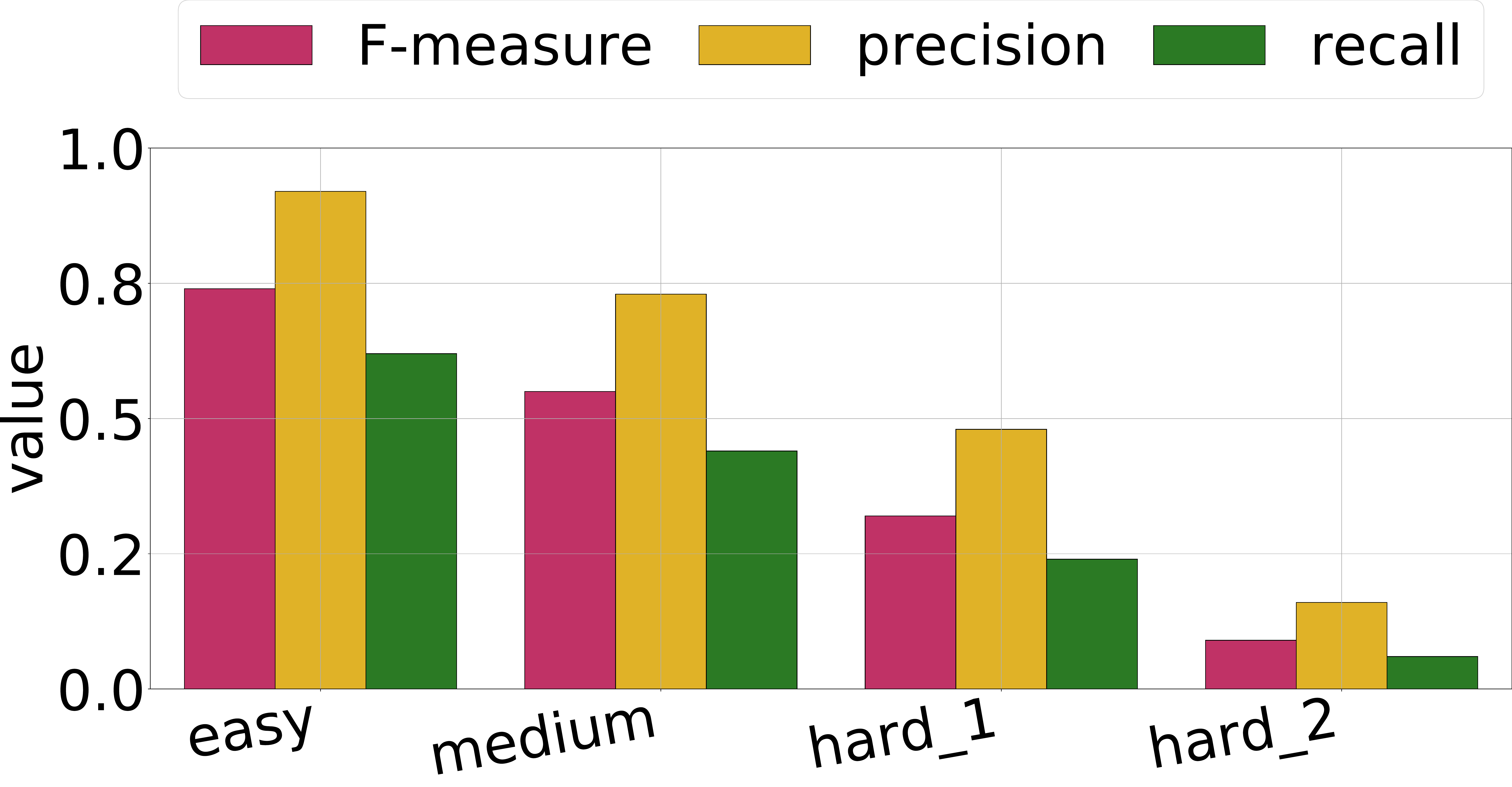}
    \caption{\flex{} results for the mediated SM task on different instances from the \texttt{camera} vertical.}
    \label{fig:flexmatcher}
\end{figure}

\myparagraph{Mediated SM} Given a mediated schema $T$, recent mediated SM approaches can use a set of sources with known attribute correspondences with $T$ -- i.e., the training sources -- to learn features of mediated attribute names and values and then use such features to predict correspondences for unseen sources -- i.e., the test sources. It is worth observing that test sources with low attribute sparsity are easier to process. It is also important that there must be enough source similarity between the test source and the training sources. %
For this task, we consider the {\flex} approach~\cite{chen2018biggorilla}. 

Figure~\ref{fig:flexmatcher} reports the performance of the \flex{} approach for four different choices for the training sources $S_{T1}$, $S_{T2}$, and test sources $S_{t}$ in the \texttt{camera} vertical. Given our manually curated mediated schema, we set $\mathcal{S}=\{S_{t}\}$. Each selected test source $t$ has different sparsity $AS(S_{t})$ as in Table~\ref{tab:flex}. For each test source $t$ we select two training sources $T1,T2$ with different average source similarity $SS=\frac{SS(S_{T1},S_{t})+SS(S_{T2},S_{t})}{2}$.

\begin{table*}
\setlength{\tabcolsep}{0.5em} %
\centering
\begin{tabular}{|l|l|l|l|l|l|}
\hline
$S_{T1}$, $S_{T2}$ & $S_{t}$ & $AS(S_{t})$ & $SS$ %
 & Difficulty \\ \hline \hline

price-hunt, & \multirow{2}{*}{mypriceindia} & \multirow{2}{*}{0.42} & \multirow{2}{*}{0.35} & \multirow{2}{*}{easy} \\ 
pricedekho & & & & \\ \hline

cambuy, & \multirow{2}{*}{buzzillions} &  \multirow{2}{*}{0.87} 
& \multirow{2}{*}{0.09} & \multirow{2}{*}{medium} \\ 
pcconnection  &  &
& & \\ \hline

cambuy, & \multirow{2}{*}{buzzillions} & \multirow{2}{*}{0.87} & \multirow{2}{*}{0.05} & \multirow{2}{*}{hard} \\
flipkart & & & & \\ \hline

cambuy, & \multirow{2}{*}{ebay} & \multirow{2}{*}{0.99} & \multirow{2}{*}{0.05} & \multirow{2}{*}{hard} \\
flipkart & & & & \\ \hline

\end{tabular}
\caption{Sources considered for our mediated SM experiment.}
\label{tab:flex}
\end{table*}

Results in Figure~\ref{fig:flexmatcher} show lower F-measure over test sources with higher attribute sparsity and lower source similarity with training sources. Note that both attribute sparsity of the test source and its source similarity with the training sources matter and solving SM over the same test source (e.g., $t=$~buzzillions) can become more difficult when selecting training sources with less source similarity.

\begin{figure}
    \centering
    \includegraphics[width=0.8\columnwidth]{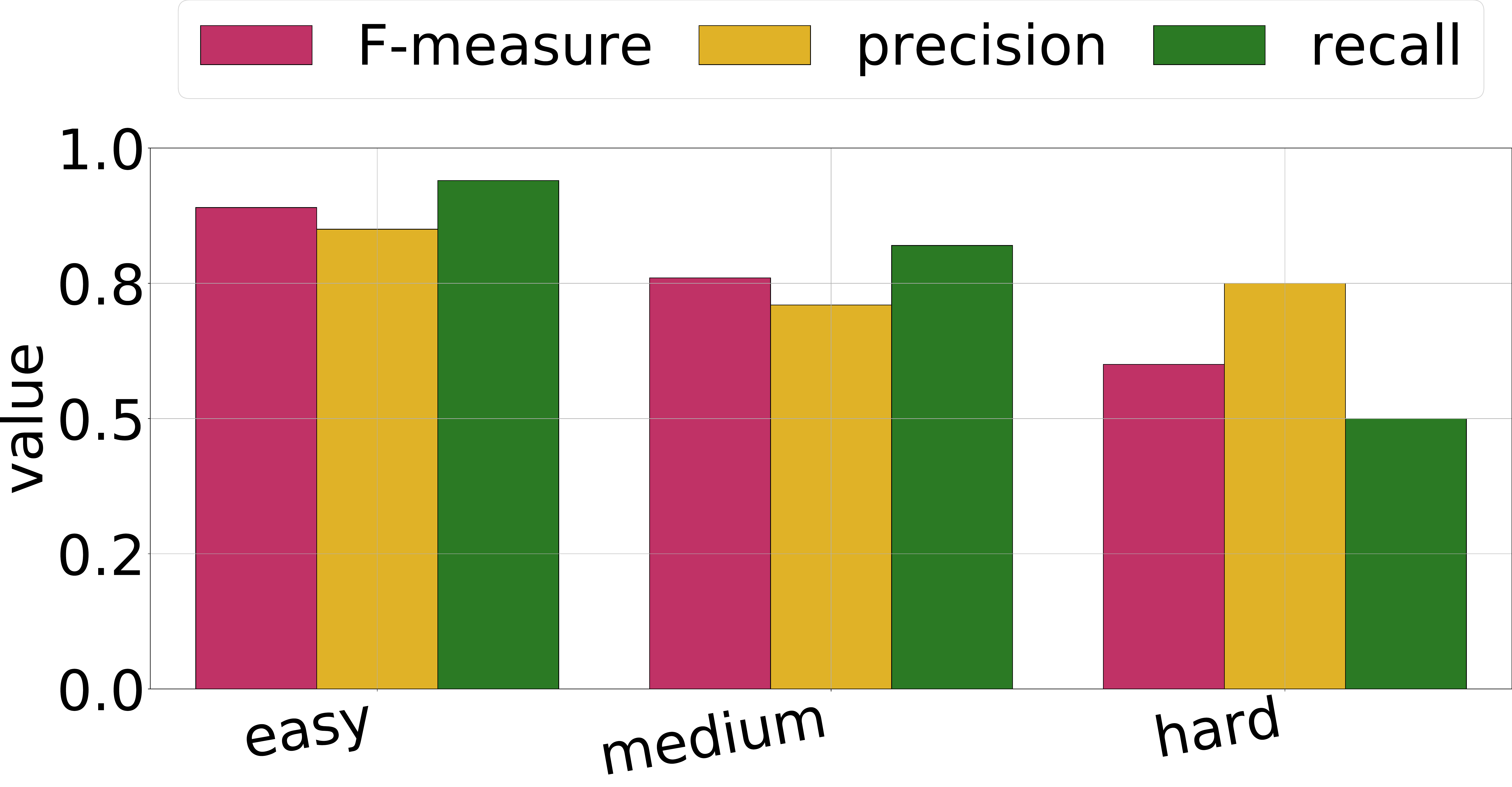}
    \caption{DeepMatcher results for the similarity-join ER task on different instances from the \texttt{camera} vertical}
    \label{fig:sb-deepmatcher}
\end{figure}

\subsubsection{Entity Resolution Scenarios}
\noindent
\textbf{Similarity-join ER} The similarity-join ER task takes as input two sources and returns the pairs of records that refer to the same entity. %
Similarity-join ER is typically easier when the sources provide as few null values as possible, which can be quantified by our attribute sparsity metric.  For this task, we consider the DeepMatcher approach~\cite{mudgal2018deep}.

Figure~\ref{fig:sb-deepmatcher} shows the performance of DeepMatcher for different choices of the input sources $S_1$ and $S_2$ from the \texttt{camera} domain. Since similarity-join ER tasks traditionally assume that the schemata of the two sources are aligned, we consider a subset of the attributes of the selected sources, and manually align them by using our SM ground truth.\footnote{Note that the $AS$ metric, for this experiment, is computed accordingly on the aligned schemata.} Considered attributes for this experiment are  \texttt{brand}, \texttt{image\_resolution} and \texttt{screen\_size}. We select pairs of sources with increasing attribute sparsity $AS(S_1 \cup S_2)$ as reported in Table~\ref{tab:sb-er-sj}. In the table, for the sake of completeness, we also report the total number of records $|S_1| + |S_2|$. It is worth saying that we can select easy/hard instances from both sides of the source size spectrum. %

\begin{table*}
\centering
\begin{tabular}{|l|l|l|l|}
\hline
$S_1,S_2$ & $AS(S_1 \cup S_2)$ & $|S_1|$ + $|S_2|$ & Difficulty \\ \hline \hline
cambuy, & \multirow{2}{*}{0.08} & \multirow{2}{*}{748} & \multirow{2}{*}{easy} \\
shopmania & & & \\
\hline
ebay, & \multirow{2}{*}{0.11} & \multirow{2}{*}{14,904} & \multirow{2}{*}{medium} \\ 
shopmania & & & \\
\hline
flipkart, & \multirow{2}{*}{0.41} & \multirow{2}{*}{338} & \multirow{2}{*}{hard} \\ 
henrys & & & \\
\hline
\end{tabular}
\caption{Sources considered for our similarity-join ER experiments.}
\label{tab:sb-er-sj}
\end{table*}

Results in Figure~\ref{fig:sb-deepmatcher} show that lower F-measure is obtained for source pairs with higher attribute sparsity. We observe that precision is reasonably high even for the harder instances: the main challenge in this scenario is indeed that of recognizing record pairs that refer to the same entity even when they have only a few non-null attributes that can be used for the purpose.  Those pairs contribute to the recall.

\begin{figure}
    \centering
    \includegraphics[width=0.8\columnwidth]{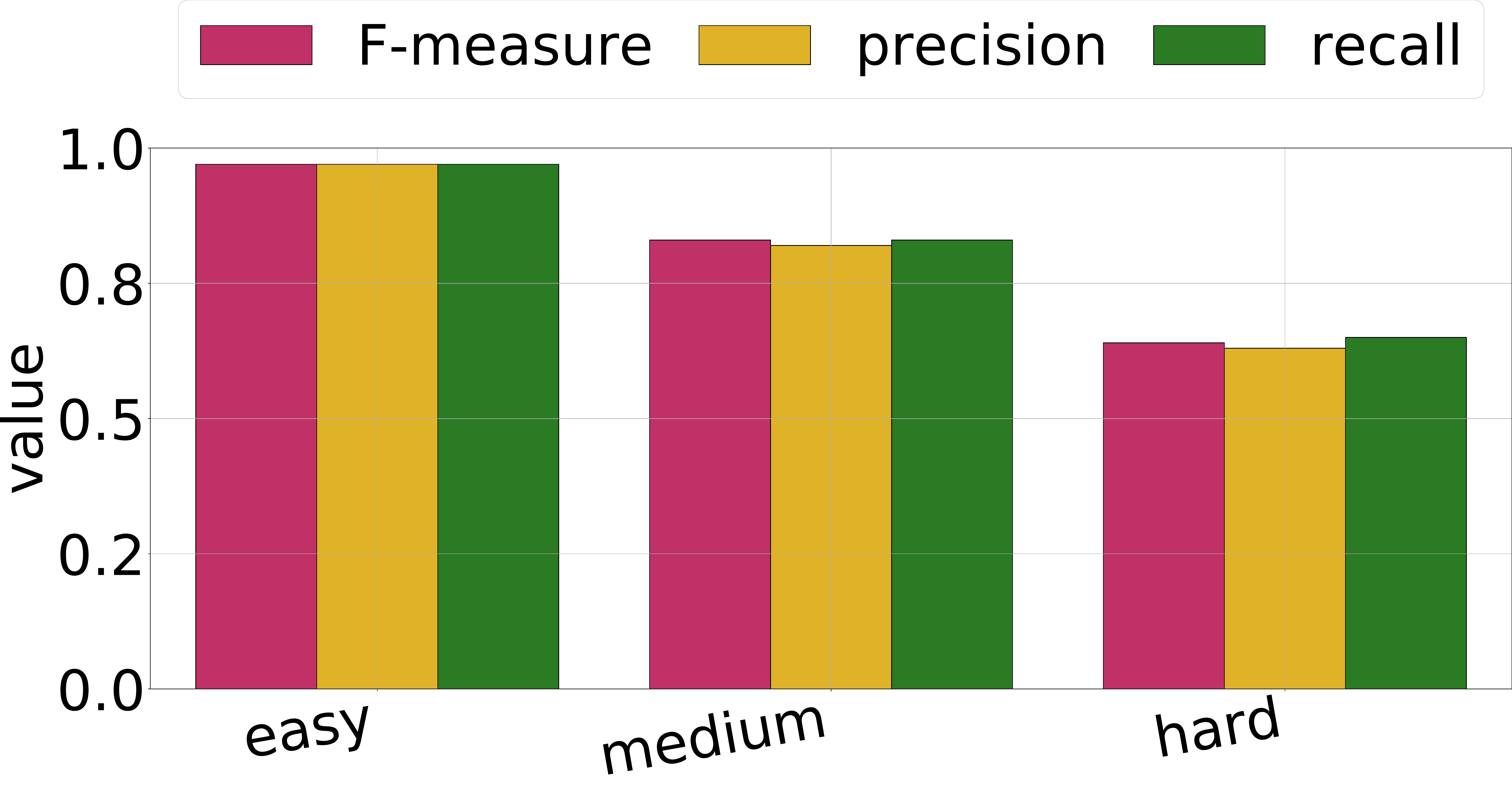}
    \caption{DeepMatcher results for the self-join ER task on different instances from the \texttt{notebook} vertical} %
    \label{fig:sb-er-self-join-results}
\end{figure}

\myparagraph{Self-join ER} Analogously to similarity-join ER approaches, self-join approaches also tend to work better when the attribute sparsity is low. In addition, a skewed cluster size distribution can make the problem harder, as there are more chances that larger clusters with many different representations of the same entity can be erroneously split, or merged with smaller clusters. %
For this task, we consider again the DeepMatcher approach~\cite{mudgal2018deep}.

Figure~\ref{fig:sb-deepmatcher} shows the performance of DeepMatcher on different subsets of 10 sources from the \texttt{notebook} collections. We choose the \texttt{notebook} vertical for this experiment as it has the largest skew in its cluster size distribution (see Figure~\ref{fig:entities-frequency}). Analogously to the similarity-join ER task, the schemata of the selected sources are manually aligned by using our SM ground truth. The considered attributes for this experiment are \texttt{brand}, \texttt{display\_size}, \texttt{cpu\_type}, \texttt{cpu\_brand}, \texttt{hdd\_capacity}, \texttt{ram\_capacity} and \texttt{cpu\_frequency}.
We select source sets with increasing attribute sparsity $AS(\bigcup_{S \in \mathcal{S}} S)$ as reported in Table~\ref{tab:sb-er-self-join-settings}. In the table, for the sake of completeness, we also report the total number of records in each source set.

\begin{table*}
\small
\centering
\begin{tabular}{|p{9cm}|l|l|l|l|c|}
\hline
$\mathcal{S}$ & $AS(\stackunder{\bigcup}{_{S \in \mathcal{S}}} S)$  & $\stackunder{\sum}{_{S \in \mathcal{S}}}|S|$ & Difficulty \\ \hline \hline
bidorbuy, livehotdeals, staples, amazon, pricequebec, vology, topendelectronic, mygofer, wallmartphoto, softwarecity &  0.16 & 7,311 & easy\\ \hline
bidorbuy, bhphotovideo, ni, overstock, livehotdeals, staples, gosale, topendelectronic, mygofer, wallmartphoto &  0.36  & 1,638 & medium \\ \hline
bidorbuy, bhphotovideo, ni, overstock, amazon, pricequebec, topendelectronic, flexshopper, wallmartphoto, softwarecity &  0.53 & 4,629 & hard \\ \hline
\end{tabular}
\caption{Sources considered for our self-join ER experiment. All sets $\mathcal{S}$ have cardinality 10.}
\label{tab:sb-er-self-join-settings}
\end{table*}

Results in Figure~\ref{fig:sb-er-self-join-results}  show lower F-measure for source groups with higher attribute sparsity, with similar decrease both in precision and recall. %

\begin{figure}
    \centering
    \includegraphics[width=0.8\columnwidth]{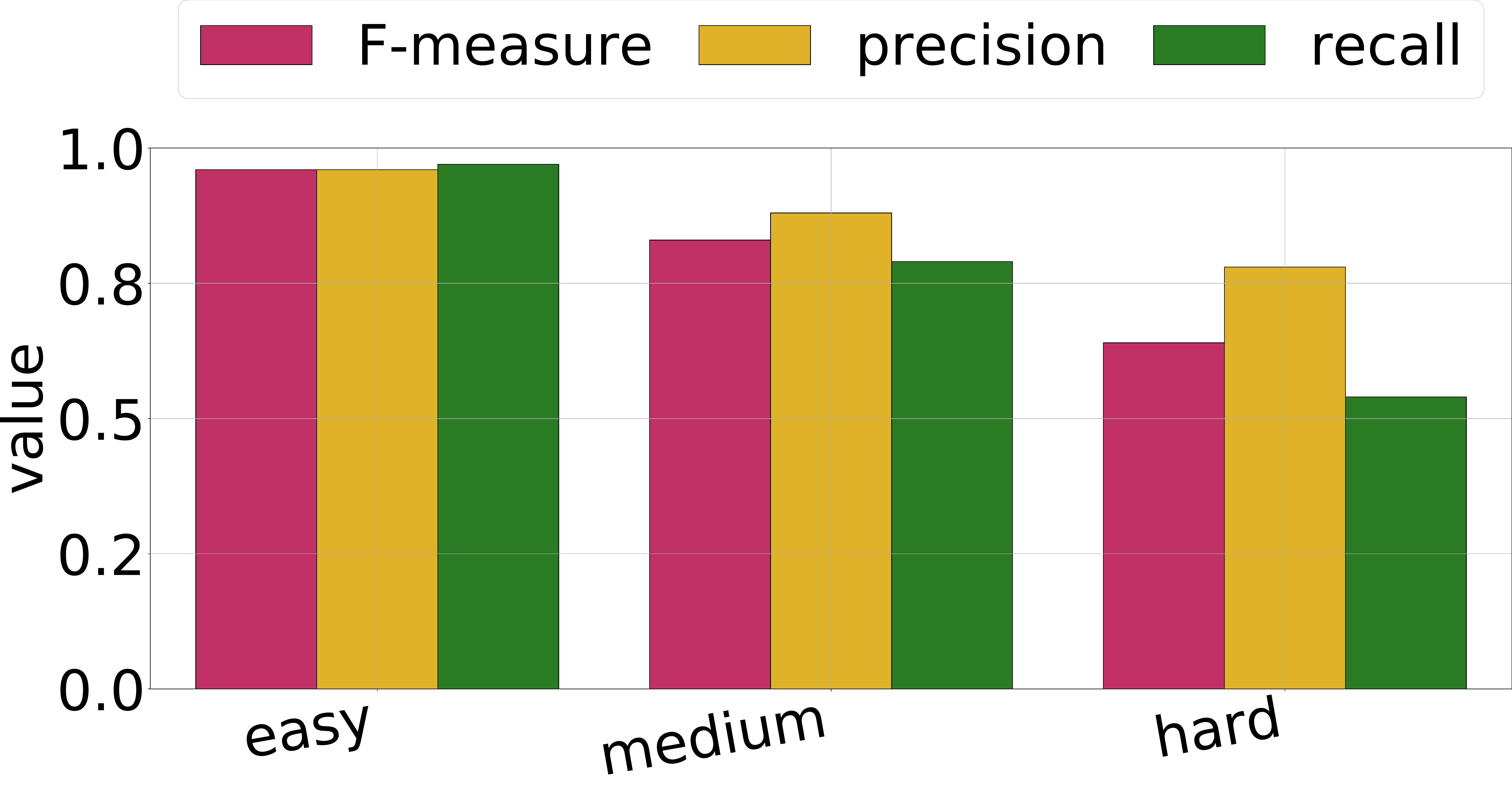}
    \caption{DeepER results for the schema-agnostic ER task on different instances from the \texttt{notebook} vertical.}
    \label{fig:deeper-sa-sj-subsets}
\end{figure}

\myparagraph{Schema-agnostic ER} Our last usage scenario consists of the same self-join ER task just discussed, but without the alignment of the schemata. The knowledge of which attribute pairs refer to the same property, available in schema-agnostic and self-join approaches, is here missing. The decision about which records ought to be matched is made solely on the basis of attribute values, concatenating all the text tokens of a record into a long descriptive paragraph. %
For this reason, the difficulty of schema-agnostic ER instances depends on the variety of the text tokens available, which can be quantified by our vocabulary size metric. For this task, we use the DeepER~\cite{ebraheem2017deeper} approach. 

Figure~\ref{fig:deeper-sa-sj-subsets} shows the performance of DeepER on different subsets of the sources in the \texttt{notebook} vertical. For each record, we concatenate the \texttt{<page title>} attribute values and the values of all other attributes in the record. We refer to the resulting attribute as \texttt{description}. We select source sets with increasing vocabulary size as reported in Table~\ref{tab:schema-agnostic-er-table}. Specifically, for each source $S \in \mathcal{S}$ we compute our $VS$ metric separately for the tokens in attribute \texttt{description} that come from the \texttt{<page title>} attribute ( $VS(S_\textnormal{title})$) and all the other tokens ( $VS(S_\textnormal{--title})$). Then we report the average $VS(S_\textnormal{title})$ and $VS(S_\textnormal{--title})$ values over all the sources in the selected set $\mathcal{S}$. In the table,  for the sake of completeness,  we also report the total number of records in each source set.

\begin{table*}
\centering
\begin{tabular}{|p{7cm}|l|l|l|l|}
\hline
$\mathcal{S}$ & $\stackunder{\textnormal{avg}} {_{S \in \mathcal{S}}} VS(S_\textnormal{title})$ & $\stackunder{\textnormal{avg}} {_{S \in \mathcal{S}}} VS(S_\textnormal{--title})$ & $\stackunder{\sum}{_{S \in \mathcal{S}}} |S|$ &  Difficulty \\ \hline
tigerdirect, bidorbuy, gosale, ebay, staples, softwarecity, mygofer, buy, thenerds, flexshopper & 1,579.3 & 19,451.0 & 11,561 & easy \\ \hline
tigerdirect, bidorbuy, gosale, wallmartphoto, staples, livehotdeals, pricequebec, softwarecity, mygofer, thenerds & 733.8 & 8,931.0 & 2,453 & medium \\ \hline
tigerdirect, bidorbuy, gosale, wallmartphoto, staples, ni, livehotdeals, bhphotovideo, topendelectronic, overstock & 458.3 & 5,135.0 & 1,621 & hard \\ \hline
\end{tabular}
\caption{Sources considered for our schema-agnostic ER experiment.}
\label{tab:schema-agnostic-er-table}
\end{table*}

Results in Figure~\ref{fig:deeper-sa-sj-subsets} show lower F-measure for source groups that have, on average, smaller vocabulary size.

\section{Data Collection}
\label{sec:coll}

In this section, we describe the data collection process behind the \alaska{} benchmark. Such process consists of the following main steps: {$(i)$}~source discovery,  {$(ii)$}~data extraction,  {$(iii)$}~data filtering, and {$(iv)$}~source filtering.

\myparagraph{Source Discovery} The first step has the goal of finding websites publishing pages with specifications for products of a vertical of interest. To this end, we used the focused crawler of the {\sf DEXTER} project~\cite{qiu2015dexter}.%
\footnote{\url{https://github.com/disheng/DEXTER}}It discovers the target websites and downloads all the product pages, i.e., pages providing detailed information about a main product.

\myparagraph{Data Extraction} The pages collected by the {\sc DEXTER} focused crawler are then processed by an ad-hoc data extraction tool -- that we call {\sc Carbonara} -- specifically developed for the \alaska{} benchmark. %
From each input page, {\sc Carbonara} extracts a product specification, that is, a set of  key-value pairs. To this end,
{\sc Carbonara} uses a classifier  which is trained to select HTML tables and lists that are likely to contain the product specification. The features used by the classifier are related to DOM properties (e.g., number of URLs in the table element, number of bold tags, etc.) and to the presence of domain keywords, i.e., terms that depends on the specific vertical. The training set, as well as the domain keywords were manually created.
The output of {\sc Carbonara} is a JSON object with all the successfully extracted key-value pairs, plus a special key-vale pairs composed by the <page title> key whose value corresponds to the HTML page title.

\myparagraph{Data Filtering} Since the extraction process is error prone, the JSON objects produced by {\sc Carbonara}  are filtered based on a heuristics, which we refer to as \textit{3-3-100} rule, with the objective of eliminating noisy attributes due to an imprecise extraction. The \textit{3-3-100} rule works as follows. %
First, key-value pairs with a key that is not present in at least $3$ pages of the same source are filtered out. Then, JSON objects with less than $3$ key-value pairs are discarded. Finally, only sources with more than $100$ JSON objects (i.e., $100$ product specifications) are gathered. 

\myparagraph{Source Filtering} Some of the sources discovered by {\sf DEXTER} can contain only \emph{copies} of specifications from other sources. The last step of the data collection process aims at eliminating these sources. To this end, we eliminate sources that are either known aggregators (such as, \url{www.shopping.com}), or country-specific versions of a larger source (such as, \url{www.ebay.ca} and \url{www.ebay.co.uk} with respect to \url{www.ebay.com}). %

\section{Deployment experiences}
\label{sec:dep}

Preliminary versions of the \alaska{} benchmark have been recently used for the 2020 SIGMOD Programming Contest\footnote{\url{http://www.inf.uniroma3.it/db/sigmod2020contest}} and for two editions of the DI2KG challenge.\footnote{\url{http://di2kg.inf.uniroma3.it}} In these experiences, we asked the participants to solve one or multiple tasks on one or more verticals. We provided participants with a subset of our manually curated ground truth for training purposes. The rest of the ground truth was kept secret and used for computing F-measure of submitted solutions. More details follow.

\myparagraph{DI2KG 2019} The DI2KG 2019 challenge was co-located with the 1st International Workshop on Challenges and Experiences from Data Integration to Knowledge Graphs (1st DI2KG workshop), held in conjunction with KDD 2019. Tasks included in this challenge were mediated SM and schema-agnostic ER. The only available vertical was \texttt{camera}. We released to participants 99 labelled records and 288 labelled attribute/mediated attribute pairs.

\myparagraph{SIGMOD 2020} The SIGMOD 2020 Programming Contest was co-located with the 2020 International ACM Conference on Management of Data. The task included in the contest was schema-agnostic ER. The only available vertical was \texttt{camera}. We released to participants 297,651 labelled record pairs (44,039 matching and 253,612 non-matching). With respect to the previous challenge, we had the chance of {$(i)$}~augmenting our ground truth by labelling approximately 800 new records; {$(ii)$}~designing a more intuitive data format for both the source data and the ground truth data.

\myparagraph{DI2KG 2020} The DI2KG 2020 challenge was co-located with the 2nd DI2KG Workshop, held in conjunction with VLDB 2020. Tasks included in this challenge were mediated SM and schema-agnostic ER. Available verticals were \texttt{monitor} and \texttt{notebook}. For \texttt{monitor} we released to the  participants 111,156 labelled record pairs (1,073 matching and 110,083 non-matching) and 135 labelled attribute/mediated attribute pairs. For \texttt{notebook}, we released to the participants 70,125 labelled record pairs (895 matching and 69,230 non-matching) and 189 labelled attribute/mediated attribute pairs. With respect to the previous challenges, we organized \emph{tracks} accounting for different task solution techniques, including supervised machine learning, unsupervised methods and methods leveraging domain knowledge, such as product catalogs publicly available on the Web.

\subsection{Future Development}

Learning from our past experiences, we plan to extend the \alaska{} benchmark working on the directions below.

\begin{itemize}
    \item Improve usability of the benchmark, to let it evolve into a full-fledged benchmarking tool;
    
    \item Include verticals with fundamentally different content, such as biological data;
    
    \item Include variants of the current verticals by removing or anonymizing targeted information in the records, such as model names of popular products; %
    
    \item Add meta-data supporting variants of the traditional F-measure, such as taking into account the difference between synonyms and homonyms in the SM tasks or between head entities (with many records) and tail entities in the ER tasks; %

    \item Collect ground truth data for other tasks in the data integration pipeline, such as Data Extraction and Data Fusion.

\end{itemize}

Finally, especially for the latter direction, we plan to reconsider the ground truth construction process in order to reduce the manual effort by re-engineering and partly automatizing the curation process.

\section{Conclusions}
\label{sec:concl}

We presented \alaska{}, a flexible benchmark for evaluating several variations of Schema Matching and Entity Resolution systems. For this purpose, we collected real-world data sets from the Web, including sources with different characteristics that provide specifications about three categories of products. We manually curated a ground both for Schema Matching and Entity Resolution tasks. We presented a profiling of the dataset under several dimensions, to show the heterogeneity of data in our collection and to allow the users to select a subset of sources that best reflect the target setting of the systems to evaluate. We finally illustrated possible usage scenarios of our benchmark, by running some representative state-of-the-art systems on different selections of sources from our dataset.

\section*{Acknowledgements}
We thank Alessandro Micarelli and Fabio Gasparetti for providing the computational resources employed in this research. 
Special thanks to Vincenzo di Cicco who contributed to implement the {\sc Carbonara} extraction system. %

\clearpage

\bibliographystyle{abbrv}
\bibliography{bibliography}

\end{document}